# Incentive-based Decentralized Routing for Connected and Autonomous Vehicles using Information Propagation

Chaojie Wang[a], Srinivas Peeta[a,b*], Jian Wang[c*]

[a]*School of Civil and Environmental Engineering, Georgia Institute of Technology, Atlanta, 30318, U.S.A.*
[b]*H. Milton Stewart School of Industrial and Systems Engineering, Georgia Institute of Technology, Atlanta, 30318, U.S.A.*
[c]*School of Transportation, Southeast University, Nanjing, 211189, China*

**Abstract**

Routing strategies under the aegis of dynamic traffic assignment have been proposed in the uliterature to optimize system performance. However, challenges have persisted in their deployment ability and effectiveness due to inherent strong assumptions on traveler behavior and availability of network-level real-time traffic information, and the high computational burden associated with computing network-wide flows in real-time. To address these gaps, this study proposes an incentive-based decentralized routing strategy to nudge the network performance closer to the system optimum for the context where all vehicles are connected and autonomous vehicles (CAVs). The strategy consists of three stages. The first stage incorporates a local route switching dynamical system to approximate the system optimal route flow in a local area based on vehicles' knowledge of local traffic information. This system is decentralized in the sense that it only updates the local route choices of vehicles in this area rather than route choices of all vehicles in the network, which circumvents the high computational burden associated with computing the flows on the entire network. The second stage optimizes the route for each CAV by considering individual heterogeneity in traveler preferences (e.g., the value of time) to maximize the utilities of all travelers in the local area. Constraints are also incorporated to ensure that these routes can achieve the approximated local system optimal flow of the first stage. The third stage leverages an expected envy-free incentive mechanism to ensure that travelers in the local area can accept the optimal routes determined in the second stage. The study analytically discusses the convergence of the local route switching dynamical system. We also show that the proposed incentive mechanism is expected individual rational and budget-balanced, which ensure that travelers are willing to participate and guarantee the balance between payments and compensations, respectively. Further, the conditions for the expected incentive compatibility of the incentive mechanism are analyzed and proved, ensuring behavioral honesty in disclosing information. Thereby, the proposed incentive-based decentralized routing strategy can enhance network performance and user satisfaction under fully connected and autonomous environments.

## 1. Introduction

The determination of origin-destination (OD) routes is a key decision for travelers, and can also entail en route decision-making to account for dynamics in network conditions, especially under information availability. Over the past three decades, dynamic traffic assignment (DTA) has been identified as the methodological engine to determine OD routes that satisfy some individual traveler objectives/constraints and/or system-wide goals in traffic networks (Peeta and Ziliaskopoulos, 2001). Advances in DTA models related to routing strategies have focused on enhancing realism and prediction accuracy. Despite these advances, challenges have persisted in a deployment context due to the complexity of the problem arising from the multiple dimensions that characterize it. First, accurate real-time information on network-level traffic conditions and/or route characteristics (such as travel time) is presumed to be available either *a priori* for the entire time horizon of interest or at the current time. While some studies (e.g., Du et al., 2012; Du et al. 2013) account for randomness in the link travel time distributions due to measurement errors or the fusion of information from multiple sources, an underlying assumption in most DTA studies is the availability of traffic data on all links of a network. While even today the presumption of ubiquitous data availability at the network level in itself is rather optimistic due to the lack of sensor coverage on all network links, the assumption of its seamless availability to all travelers in real-time is unrealistic in a deployment context. Hence, the availability of massive amounts of information required for DTA should not be presumed to be trivial or seamless, especially when the reliability of the predictions depends heavily on the accuracy and timeliness of the information.

Second, the heterogeneity in traveler characteristics in the context of making routing decisions is not modeled satisfactorily. For example, the use of user classes (Peeta and Mahmassani, 1995a) masks differences in responses among the individuals of a user class. Further, there is randomness in the fractions of different user classes in the traffic stream with time. Hence, while the consideration of multiple user classes may be useful in identifying routing strategies for planning purposes, the accuracy of state prediction is lacking in a real-time deployment context. This motivates the consideration of real-time network traffic management strategies that directly capture individual heterogeneity related to traveler characteristics.

Third, computational tractability has always been a challenge for the real-time deployment of DTA models, especially for real-world urban networks. Due to the complexity of the DTA problem, analytical solutions with adequate levels of modeling realism are lacking in a deployment context. Even for models with simplified assumptions, the computational time of analytical solutions escalates rapidly with network size. Hence, simulation-based algorithms (Peeta and Mahmassani, 1995b) have been adopted in real-world implementations. However, the computational burden for deploying these algorithms can be significant when determining solutions in a centralized manner, leading to mitigation strategies such as the use of mesoscopic traffic simulators, and the deployment of distributed/decentralized control-based solutions (Hawas and Mahmassani, 1997; Pavlis and Papageorgiou, 1998).

Over the years, various studies have leveraged DTA models to additionally explore mechanisms to influence travelers' routing decisions so as to push the network performance closer to the system optimal solution. Brueckner et al. (2001) recommend demand-side solutions to mitigate traffic congestion rather than the use of expensive supply-side mechanisms such as the construction of new lanes or roads. In this context, incentive-based approaches are gaining attention as mechanisms to leverage the heterogeneity in individual behavior to enhance system performance. However, many practical incentive-based approaches (Merugu et al., 2009)) to influence individual travelers require monetary investment from one or more stakeholders. Further, existing incentive mechanisms primarily aim to influence macro travel decisions like travel mode choice (Kazhamiakin et al., 2015) and departure time choice. However, incentive mechanisms to influence micro travel decisions (such as real-time route choice), for which opportunities exist frequently in real-world networks, are rare. Partly, this can be attributed to safety concerns associated with providing real-time incentives, arising from distraction, cognitive burden, and limited attention span of drivers due to the inherently multitasking environment of congested traffic networks. Hence, such incentives have only been proposed in the routing context for pre-trip route choice (Hu et al., 2015). The problem is also challenging because micro travel decisions occur in network traffic conditions characterized by dynamics and randomness. Further, centralized incentive mechanisms that would potentially target individuals for route-related incentives in a coordinated manner require understanding/knowledge of their behavioral characteristics, which is challenging in a deployment context. Moreover, there is a lack of theoretical analysis on the participation willingness and behavioral honesty in existing incentive mechanisms, which are both critical concerns in practical implementation.

The emerging disruptive and transformative technologies of automation and connectivity provide several enablers to foster the development of a new generation of incentive mechanisms that target micro travel decisions by individuals to enhance system performance. A connected transportation environment consisting of vehicle-to-vehicle (V2V) and vehicle-to-infrastructure (V2I) communications enables vehicles to have access to real-time traffic information by leveraging vehicles' ability to communicate their travel experience (for example, link travel time, location, speed, etc.) to each other (through V2V) and the system (through V2I), thereby obviating the need for ubiquitous network-wide installation of sensors (e.g., Bagloee et al., 2017). Some studies (Kim et al., 2016; Kim et al., 2017; Wang et al., 2018; Wang et al., 2019; Kim and Peeta, 2019) have proposed models related to information propagation to alleviate the computational burden and scalability issues with centralized information communication-based strategies. However, information propagation suffers from accumulated communication delay. For the timeliness and accuracy of the traffic information, information propagation is more likely to be meaningfully applied in local information updating.

While the leveraging of connected technologies through V2V communications can enable seamless access to local traffic information for vehicles, and mitigate computational burden issues through the fostering of decentralized routing strategies, technologies associated with automation and autonomy provide additional capabilities to address the other DTA deployment limitations that were identified earlier. Hence, while V2V-based information propagation enables efficient communications between vehicles, automation allows vehicles to react and respond instantly to the information received. Thereby, it is possible to leverage complex incentive mechanisms, which factor the individual heterogeneity, to develop decentralized incentive-based routing strategies for CAVs.

This study proposes a new incentive-based routing strategy to leverage connectivity and automation technologies to address the aforementioned gaps related to existing routing strategies using DTA models. First, vehicular communication-based information propagation enables local information availability without incurring computational burden for a central server. Second, the proposed routing strategy fully factors the individual heterogeneity in the behaviors of vehicles through individual route evaluation functions rather than modeling user class behavior like conventional DTA models to estimate the likely route decisions of vehicles. Hence, the routing decisions are more representative as the characteristics and preferences of each vehicle (traveler) are fully reflected through individual evaluation functions. Third, the decentralized nature of the routing strategy makes it both flexible and computational tractable compared with current predetermined or centralized incentive schemes. Also, the cooperation enabled through the decentralized route assignment model leads to better system performance, unlike under V2V communication-enabled uncoordinated routing decisions by individual vehicles. Overall, the proposed incentive-based decentralized routing strategy for CAVs can enhance system performance and promote user satisfaction under realistic information availability assumptions while factoring individual heterogeneity and being computationally tractable.

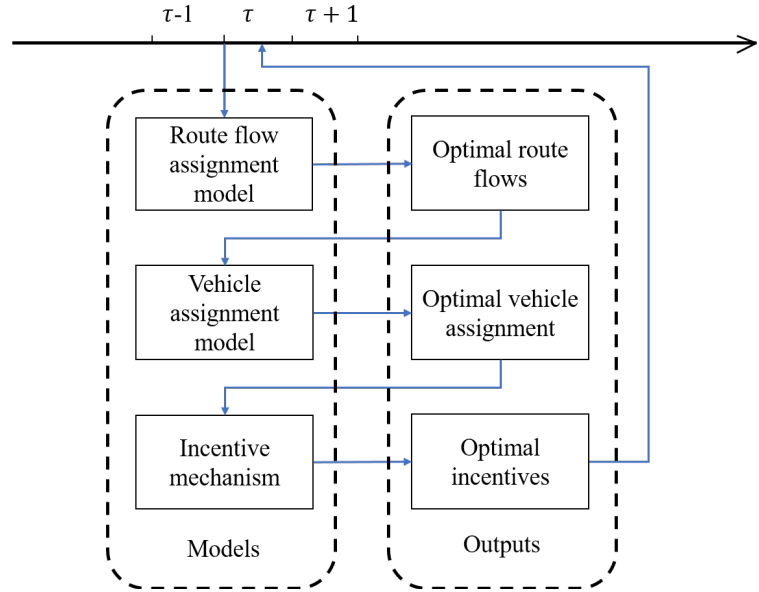

Fig. 1. Flow chart of the proposed decentralized routing strategy.

Fig. 1 illustrates the conceptual flow chart of the routing strategy. The time horizon of interest is divided into small time intervals, within which only local traffic information is updated in real-time, while that in the rest of the network is considered unchanged. In each time interval, vehicle groups generated according to their origin-destination (OD) pair information will calculate the optimal routes that cooperatively enhance system performance, assign vehicles to corresponding routes, and develop optimal incentives that nudge agents† to follow the route suggestions. At the beginning of the next interval, due to the updated boundary of the local area and possibly updated information related to further areas, the optimal routes will likely change. Therefore, the routing strategy will be executed iteratively.

At the beginning of each time interval $\tau$, vehicles of the vehicle group corresponding to an OD pair will start a decentralized local route flow assignment protocol. Following the protocol, the vehicle group will update the route flows according to the local traffic information and disseminate them using local information propagation to update the local traffic information. This procedure is repeated until a stable state is reached. We will show that the route flows of the stable state are optimal route flows, which can achieve good network-level system performance. The optimal route flows will be the inputs for the vehicle route assignment model, which assigns vehicles in the group to the different routes to achieve the optimal route flows by considering individual route evaluation functions. The individual route evaluation function captures the personalized characteristics and preferences over different routes, which enables the vehicle route assignment model to maximize the group utility with the full consideration of individual heterogeneity. The optimal vehicle route assignments are the inputs to the incentive mechanism, which determines the incentives for agents in the vehicle group that ensure that vehicle route assignments are fair/satisfactory for each agent. We will show that for some specific situations, optimal incentives developed by the mechanism are:

---

† Following the tradition in economics, we will use the term agent instead of vehicle/driver/traveler when describing utility-related models.

(i) budget-balanced, which indicates no monetary investment is required in the long term; (ii) expected individual rational, which implies that agents are willing to participate repeatedly; and (iii) expected incentive compatible, implying agents will behave honestly for long-term benefit under the proposed mechanism. Therefore, the optimal incentives will nudge drivers to follow the optimal vehicle route assignment, thus achieving the optimal route flows in the time interval $\tau$. At the end of interval $\tau$, since the boundary of the local area changes and the traffic information related to further areas may have been updated, this strategy is repeated in the next time interval $\tau + 1$ with the updated conditions.

This paper is organized as follows. Section 2 describes the problem and justifies the iterative local routing strategy. In Section 3, we present the decentralized local route assignment model and prove the system properties it can achieve. The vehicle route assignment model is discussed in Section 4. In Section 5, we present the fair incentive mechanism, and illustrate that it is budget-balanced, expected individual rational, and expected incentive compatible under specific situations. We conclude the paper in Section 6 with some comments.

## 2. Problem Description

Consider a dynamic traffic network $G(N, E)$, where $N$ is the node set, and $E$ is the directed link set. This study seeks to assign time-dependent routes to vehicles by considering heterogeneous individual preferences to achieve objectives at both the system and individual levels. The time horizon of interest is divided into small equal time intervals. In each time interval $\tau$, the route updating scheme contains three stages. In the first stage, a decentralized dynamic system is used to determine the approximated system optimal route flows in the local area based on vehicles' knowledge of the local traffic information. In the second stage, each vehicle is assigned a route based on its self-reported individual preferences of the routes to achieve the approximated optimal route flows obtained in the first stage. The last stage incorporates an incentive mechanism to compensate or charge the individuals based on individual heterogeneities to ensure everyone can accept the assigned routes. Hence, to ensure that the mechanism is budget-balanced to preclude the need for external funding, the benefiting vehicles compensate the sacrificing vehicles.

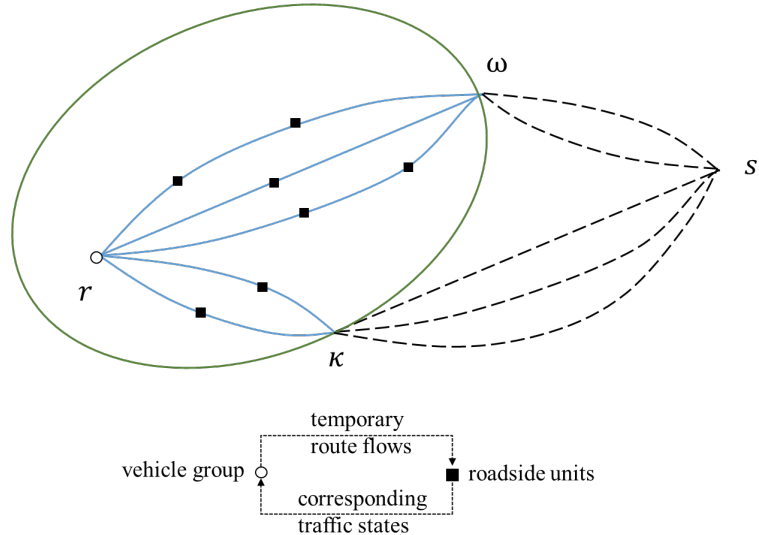

Fig. 2. Real-time traffic information availability range

Fig. 2 shows the context of this study. Suppose at the beginning of time interval $\tau$, a stream of vehicles just arrive at the node $r$. As the traffic flows can change dynamically, the traffic conditions far away from node $r$ at the current time may not be identical to the conditions that vehicles at node $r$ will experience in the future at those locations. Thereby, this study assumes only the real-time traffic information in the local area (area within the green circle) will be provided to these vehicles at node $r$ through infrastructure-to-vehicle communications. The traffic information outside of the local area is not updated for vehicles at node $r$ during the time interval $\tau$. Specifically, during the time interval $\tau$, the vehicle group (indicated by the little circle in Fig. 2) will disseminate their travel information (number of vehicles, destinations, individual evaluation functions, temporary routes, etc.) to the roadside units within the local area (black squares within the green circle in Fig. 2). The roadside units will then predict the future traffic states based on the received information on temporary route flows and then distribute the updated route and payment/compensation plans to all of these vehicles. In Fig. 2, the local area related to the origin node $r$ is included in the green circle, while the destination node $s$ is outside the local area. $\omega$ and $\kappa$ are nodes close to the boundary of the local area. We will label $(r, \omega)$ and $(r, \kappa)$ the local OD pairs and $(r, s)$ the global OD pair. Let $d$ be the range that the local area information covers. The value of $d$ depends on the information propagation speed, roadside unit coverage, etc. Further, the vehicles just arriving at node $r$ at the beginning of time interval $\tau$ also send information to the roadside units, such as the chosen route, the origins, the destinations, and travelers' characteristics (e.g., the value of time), etc. The roadside units will then determine the optimal route for each vehicle in this stream based on real-time traffic information and projection of future traffic conditions based on each vehicle's information. It should be noted that new vehicles may pass node $r$ during the time when the optimal schemes are being computed; these vehicles will not be included in the global OD demand at the beginning of $\tau$ to determine their optimal route choices because they may not be able to change their route choices accordingly.

Let $R \times S$ be the set of OD pairs in the network, and $(r, s) \in R \times S$ denotes a global OD pair. Let $\Omega_{rs}^d = \{\omega, \kappa \cdots \}$ be the set of nodes close to the boundary that is used by routes of the global OD pair $(r, s)$. For simplicity, we will denote $\Omega_{rs}^d$ as the set of local destinations. Denote $P^\omega$ as the set of local routes for local OD pair $(r, \omega)$. As the local routes are part of the corresponding global routes, the number of routes in the local OD pairs is less than that in the global OD pair. For example, in Fig. 2, the global route set for global OD pair $(r, s)$ contains 12 routes, while the local route sets for local OD pairs $(r, \omega)$ and $(r, \kappa)$ only contains 5 routes. Note that the difference would be more significant if a smaller part of the trip is within the green circle because the number of routes will increase rapidly with the increase in the number of en route bifurcations, which implies that our local route assignment dynamical system consists of much fewer state variables compared with existing approaches. Also, only vehicles in the local area are considered in the route switching dynamical system to compute the local system optimal flows. Thereby, it is highly efficient computationally by avoiding the inclusion of all vehicles in the network for computing the network-level system optimal flows.

It should be noted that the length of time interval $\tau$ determines the interval for updating the vehicles' routes; that is, vehicles will update their routes in each time interval $\tau$. To enhance traffic performance, improved routes are provided to vehicles based on real-time traffic conditions. Thereby, the value of $\tau$ should be such that vehicles can update their routes before reaching the local destination using the method proposed above.

## 3. Route Flow Assignment

This section describes the details of the route switching dynamical system to approximate the system optimal route flow in the local area for each time interval $\tau$. Denote $x_{\omega i}^{\tau rs}(t)$ as the flow of route $i$ for local OD pair $(r, \omega)$ at time $t$ in the dynamical system in time interval $\tau$ for the global OD pair $(r, s)$. $\boldsymbol{x}^\tau(t)$ is the vector of all $x_{\omega i}^{\tau rs}(t)$. Since we focus on analyzing the flow in one time interval, for simplicity, the time interval indicator will be removed, i.e., we will use $x_{\omega i}^{rs}$ and $\boldsymbol{x}$ to denote $x_{\omega i}^{rs}(t)$ and $\boldsymbol{x}^\tau(t)$ respectively, hereafter.

As discussed earlier, the sum of local OD demands that are associated with the same global OD pair (i.e., $(r, \omega)$ and $(r, \kappa)$ in Fig. 2 are both associated with the global OD pair $(r, s)$) does not change over the time unit $t$ in the dynamical system. However, different from general static traffic assignment problems, the local OD demands do not remain unchanged in the process as vehicles heading to $s$ can switch between $(r, \omega)$ and $(r, \kappa)$. Let $C_i^\omega \equiv C_i^\omega(t)$ be the marginal travel time of the local route $i$ for local OD pair $(r, \omega)$. When the system reaches a local system optimal (LSO) solution, the marginal travel time of used local routes for the same local OD pair is the same. However, differences exist between the marginal travel time of used local routes for different local OD pairs because the marginal travel times of used routes for OD pairs beyond the local area are assumed to be fixed. For example, in Fig. 2, the marginal travel times of used local routes for OD pair $(r, \omega)$ are identical at LSO, as are those for local OD pair $(r, \kappa)$. But the marginal travel times of local routes for OD pair $(r, \omega)$ are different from those of OD pair $(r, \kappa)$. The difference is equal to the difference of the fixed marginal travel time of routes for OD pairs $(\kappa, s)$ and $(\omega, s)$.

In other words, at the LSO state, global routes of a global OD pair have the same marginal travel time given the limited information outside the local area, which makes the global system state a system optimal (SO) one. We assume that the average marginal travel time difference between $(\kappa, s)$ and $(\omega, s)$, $\Delta_{\omega \kappa}$, is known at the beginning of $\tau$, and is not updated during the solution procedure of the dynamical system since it is related to road segments outside the local area.

Note that the traffic information of the network beyond the local area network cannot be predicted accurately due to the potential for high variance. Therefore, it is reasonable to pursue an approximated local system optimal (ALSO) solution rather than a strict LSO. Let $\delta_{rs}$, $\delta_{rs} \geq 0$, be the tolerance of the marginal travel time of used routes for global

OD pair $(r,s)$, i.e., ALSO is reached if the difference in the marginal travel times of two arbitrary used routes for global OD pair $(r,s)$ is within $[-\delta_{rs}, \delta_{rs}]$.

*3.1. Route Switching Model*

This section proposes a route-switching dynamical model to obtain the route flows for the ALSO by leveraging the work of Smith (1984). If we introduce the rectifier function:

$$\Gamma(x) = \max(x, 0), \tag{1}$$

we can represent the route switching logic for local route $i$ of local OD pair $(r, \omega)$ as

$$x^{rs}_{\omega(j\to i)} = \begin{cases} \Gamma(C^\omega_j - C^\omega_i - \delta_{rs})x^{rs}_{\omega j}, & \text{if } j \in P^\omega \text{ and } j \neq i; \\ \Gamma(C^\kappa_j - C^\omega_i - \Delta_{\kappa\omega} - \delta_{rs})x^{rs}_{\kappa j}, & \text{if } j \in P^\kappa \text{ and } \kappa \in \Omega^d_{rs}\setminus\{\omega\}. \end{cases} \tag{2}$$

Equation (2) describes the route flow switching rate from route $j$ to route $i$ under different conditions of $j$. The logic is similar to Smith's model other than the incorporation of $\Delta_{\kappa\omega}$ and $\delta_{rs}$ to factor the aforementioned properties of the local route assignment problem. Note that $\Delta_{\omega\omega} = 0$. According to Equation (2), the total change rate of the flow on local route $i$ for OD pair $(r, \omega)$ is

$$\dot{x}^{rs}_{\omega i} = \sum_{\kappa \in \Omega^d_{rs}} \sum_{j \in P^\kappa \text{ and } j \neq i} \left[ \Gamma(C^\kappa_j - C^\omega_i - \Delta_{\kappa\omega} - \delta_{rs})x^{rs}_{\kappa j} - \Gamma(C^\omega_i - C^\kappa_j - \Delta_{\omega\kappa} - \delta_{rs})x^{rs}_{\omega i} \right] \tag{3}$$

Considering all global OD pairs $(r,s) \in R \times S$, the dynamical system that describes the local route switching process can be represented as:

$$\dot{x} = \Phi(x) \cdot x = \begin{bmatrix} \Phi_{r_1 s_1}(x) & 0 & \cdots & \cdots & 0 \\ 0 & \Phi_{r_1 s_2}(x) & 0 & \cdots & \cdots \\ \cdots & 0 & \cdots & 0 & \cdots \\ \cdots & \cdots & 0 & \Phi_{r_2 s_1}(x) & 0 \\ 0 & \cdots & \cdots & 0 & \cdots \end{bmatrix} \cdot x, \tag{4a}$$

where

$$\Phi_{rs}(x) = \begin{bmatrix} \Psi^{rs}_{\omega_1 \omega_1} & \Psi^{rs}_{\omega_1 \omega_2} & \cdots \\ \Psi^{rs}_{\omega_2 \omega_1} & \Psi^{rs}_{\omega_2 \omega_2} & \cdots \\ \cdots & \cdots & \ddots \end{bmatrix} \tag{4b}$$

is the coefficient matrix related to the route switching process of the global OD $(r,s)$,

$$\Psi^{rs}_{\omega\omega} = \begin{bmatrix} \Psi^{rs}_{\omega\omega}[1,1] & \Gamma(C^\omega_1 - C^\omega_2 - \delta_{rs}) & \cdots & \Gamma(C^\omega_1 - C^\omega_i - \delta_{rs}) & \cdots \\ \Gamma(C^\omega_2 - C^\omega_1 - \delta_{rs}) & \Psi^{rs}_{\omega\omega}[2,2] & \cdots & \Gamma(C^\omega_2 - C^\omega_i - \delta_{rs}) & \cdots \\ \cdots & \cdots & \ddots & \cdots & \cdots \\ \Gamma(C^\omega_i - C^\omega_1 - \delta_{rs}) & \Gamma(C^\omega_i - C^\omega_2 - \delta_{rs}) & \cdots & \Psi^{rs}_{\omega\omega}[i,i] & \cdots \\ \cdots & \cdots & \cdots & \cdots & \cdots \end{bmatrix} \tag{4c}$$

is the coefficient matrix related to the flow switching within the local route set $P^\omega$,

$$\Psi^{rs}_{\omega\omega}[i,i] = -\sum_{j\in P^\omega \setminus \{i\}} \Gamma(C^\omega_j - C^\omega_i - \delta_{rs}) - \sum_{\kappa \in \Omega^d_{rs}} \sum_{j\in P^\kappa} \Gamma(C^\kappa_j - C^\omega_i - \Delta_{\kappa\omega} - \delta_{rs}) \tag{4d}$$

and

$$\Psi^{rs}_{\omega\kappa} = \begin{bmatrix} \Gamma(C^\omega_1 - C^\kappa_1 - \Delta_{\omega\kappa} - \delta_{rs}) & \Gamma(C^\omega_1 - C^\omega_2 - \Delta_{\omega\kappa} - \delta_{rs}) & \cdots & \Gamma(C^\omega_1 - C^\omega_i - \Delta_{\omega\kappa} - \delta_{rs}) & \cdots \\ \Gamma(C^\omega_2 - C^\omega_1 - \Delta_{\omega\kappa} - \delta_{rs}) & \Gamma(C^\omega_2 - C^\kappa_2 - \Delta_{\omega\kappa} - \delta_{rs}) & \cdots & \Gamma(C^\omega_2 - C^\omega_i - \Delta_{\omega\kappa} - \delta_{rs}) & \cdots \\ \cdots & \cdots & \ddots & \cdots & \cdots \\ \Gamma(C^\omega_i - C^\omega_1 - \Delta_{\omega\kappa} - \delta_{rs}) & \Gamma(C^\omega_i - C^\omega_2 - \Delta_{\omega\kappa} - \delta_{rs}) & \cdots & \Gamma(C^\omega_i - C^\kappa_i - \Delta_{\omega\kappa} - \delta_{rs}) & \cdots \\ \cdots & \cdots & \cdots & \cdots & \cdots \end{bmatrix} \tag{4e}$$

is the coefficient matrix related to flow switching from local routes in $P^\kappa$ to local routes in $P^\omega$.

*3.2. System Properties*

This section presents two important properties of the proposed dynamical system (4): flow non-negativity and flow conservation. The former property ensures that the route flow is always non-negative during the switching process, and the latter property ensures that the total demand of all local OD pairs equals the demand of the corresponding global OD pairs.

**Theorem 1.** *If the initial flow of all routes is non-negative, the local route flows determined by the dynamical system in Equation (4) are always non-negative.*

*Proof.* Suppose there exists a route whose flow becomes negative in the switching process. As the dynamical system is continuous, the route flow must reach 0 before becoming negative. Without loss of generality, let the route be $i$, i.e., $x_{\omega i}^{rs} = 0$. As all other routes are non-negative, we have $x_{\kappa j}^{rs} \geq 0, i \in P^{\omega}, j \in P^{\kappa}, j \neq i, \omega, \kappa \in \Omega_{rs}^{d}, (r,s) \in R \times S$. According to Equation (3),

$$\dot{x}_{\omega i}^{rs} = \sum_{\kappa \in \Omega_{rs}^{d}} \sum_{j \in P^{\kappa} \text{ and } j \neq i} \Gamma(C_j^{\kappa} - C_i^{\omega} - \Delta_{\kappa\omega} - \delta_{rs}) x_{\kappa j}^{rs} \geq 0 \tag{5}$$

Eq. (5) indicates that $x_{\omega i}^{rs}$ will never be negative if the initial states of all route flows are non-negative. □

As discussed earlier, the sum of local route flows for the same local OD pair is not conserved. However, the summation of the demand of all local OD pairs belonging to the same global OD pair is conserved. It should be equal to the OD demand for the corresponding global OD pair. The following theorem discusses this fact.

**Theorem 2.** *Let $D_{rs}^{\tau}$ be the demand at node $r$ destined to node $s$ at the beginning of $\tau$. Then, $\sum_{\omega \in \Omega_{rs}^{d}} \sum_{i \in P^{\omega}} x_{\omega i}^{rs} = D_{rs}^{\tau}$ always holds for the dynamical system defined in Equation (4).*

*Proof.* From Equation (4), we have:

$$\sum_{\omega \in \Omega_{rs}^{d}} \sum_{i \in P^{\omega}} \dot{x}_{\omega i}^{rs} = \sum_{\omega \in \Omega_{rs}^{d}} \sum_{i \in P^{\omega}} \sum_{\kappa \in \Omega_{rs}^{d}} \sum_{j \in P^{\kappa} \text{ and } j \neq i} \Gamma(C_j^{\kappa} - C_i^{\omega} - \Delta_{\kappa\omega} - \delta_{rs}) x_{\kappa j}^{rs} \\ - \sum_{\omega \in \Omega_{rs}^{d}} \sum_{i \in P^{\omega}} \sum_{\kappa \in \Omega_{rs}^{d}} \sum_{j \in P^{\kappa} \text{ and } j \neq i} \Gamma(C_i^{\omega} - C_j^{\kappa} - \Delta_{\omega\kappa} - \delta_{rs}) x_{\omega i}^{rs} = 0. \tag{6}$$

Therefore, $\sum_{\omega \in \Omega_{rs}^{d}} \sum_{i \in P^{\omega}} x_{\omega i}^{rs} = D_{rs}^{\tau}$ always holds. □

Theorems 1 and 2 imply that the feasible state vector set is positively invariant under the dynamical system.

*3.3. Convergence Analysis*

We show that for arbitrary initial route flows, the route flows determined by Equation (4) will always converge to the ALSO solution.

**Theorem 3.** *(LaSalle's theorem). Let $\Omega \subset D$ be a compact set that is positively invariant with respect to the autonomous system $\dot{x} = f(x)$. Let $V: D \to R$ be a continuously differentiable function such that $\dot{V} \leq 0$ in $\Omega$. Let $E$ be the set of all points in $\Omega$ where $\dot{V}(x) = 0$. Let $M$ be the largest invariant set in $E$. Then, every solution starting in $\Omega$ approaches $M$ as $t \to \infty$ (Khalil, 1996).*

**Lemma 1.** *The dynamical system defined by Equation (4) is an autonomous (time-invariant) system.*

*Proof.* As discussed earlier, the marginal travel time $C_i^{\omega}$ is obtained through the updated real-time local route flows, which are the state variables $\boldsymbol{x}$. Therefore, the ODEs in Equation (4) do not explicitly depend on $t$. That is to say, the dynamical system is a time-invariant system. □

We now investigate the candidate Lyapunov function:

$$V(\boldsymbol{x}) = \sum_{(r,s) \in R \times S} \sum_{\omega \in \Omega_{rs}^{d}} \sum_{\kappa \in \Omega_{rs}^{d}} \sum_{i \in P^{\omega}} \sum_{j \in P^{\kappa}} \Gamma^{2}(C_i^{\omega} - C_j^{\kappa} - \Delta_{\omega\kappa} - \delta_{rs}) x_{\omega i}^{rs} \tag{7}$$

which is a differentiable distance measure from $\boldsymbol{x}$ to ALSO.

**Lemma 2.** *If the link travel time on each link is continuously differentiable and non-decreasing, $\dot{V} \leq 0$ for the dynamical system defined in Equation (4) and the equality holds only when $\dot{\boldsymbol{x}} = 0$.*

*Proof.* Note that

$$\frac{\partial V}{\partial x_{\omega i}^{rs}} = \sum_{(r',s') \in R \times S} \sum_{\omega' \in \Omega_{r's'}^{d}} \sum_{\kappa' \in \Omega_{r's'}^{d}} \sum_{i' \in P^{\omega'}} \sum_{j' \in P^{\kappa'}} 2\Gamma\left(C_{i'}^{\omega'} - C_{j'}^{\kappa'} - \Delta_{\omega'\kappa'} - \delta_{r's'}\right) x_{\omega'i'}^{r's'} \left(\frac{\partial C_{i'}^{\omega'}}{\partial x_{\omega i}^{rs}} - \frac{\partial C_{i'}^{\omega'}}{\partial x_{\omega i}^{rs}}\right) \\ + \sum_{\kappa \in \Omega_{rs}^{d}} \sum_{j \in P^{\kappa}} \Gamma^{2}(C_i^{\omega} - C_j^{\kappa} - \Delta_{\omega\kappa} - \delta_{rs}) \tag{8}$$

Denote $\boldsymbol{J}$ as the Jacobian matrix of the vector of the marginal travel time function. If we use $[i, \omega, r, s]$ as the index for the element corresponding to the index of $x_{\omega i}^{rs}$ in $\boldsymbol{x}$, we have:

$$\left((\dot{\boldsymbol{x}}^{T})\boldsymbol{J}\right)[i, \omega, r, s] \tag{9}$$

$$= \sum_{(r',s')\in R\times S} \sum_{\omega'\in\Omega^d_{r's'}} \sum_{i'\in P^{\omega'}} \dot{x}^{r's'}_{\omega'i'} \frac{\partial C^{\omega'}_{i'}}{\partial x^{rs}_{\omega i}}$$

$$= \sum_{(r',s')\in R\times S} \sum_{\omega'\in\Omega^d_{r's'}} \sum_{\kappa'\in\Omega^d_{r's'}} \sum_{i'\in P^{\omega'}} \sum_{j'\in P^{\kappa'}} \Gamma\left(C^{\kappa'}_{j'} - C^{\omega'}_{i'} - \Delta_{\kappa'\omega'} - \delta_{r's'}\right) x^{r's'}_{\kappa'j'} \frac{\partial C^{\omega'}_{i'}}{\partial x^{rs}_{\omega i}}$$

$$- \sum_{(r',s')\in R\times S} \sum_{\omega'\in\Omega^d_{r's'}} \sum_{\kappa'\in\Omega^d_{r's'}} \sum_{i'\in P^{\omega'}} \sum_{j'\in P^{\kappa'}} \Gamma\left(C^{\omega'}_{i'} - C^{\kappa'}_{j'} - \Delta_{\omega'\kappa'} - \delta_{r's'}\right) x^{r's'}_{\omega'i'} \frac{\partial C^{\omega'}_{i'}}{\partial x^{rs}_{\omega i}}$$

$$= -\sum_{(r',s')\in R\times S} \sum_{\omega'\in\Omega^d_{r's'}} \sum_{\kappa'\in\Omega^d_{r's'}} \sum_{i'\in P^{\omega'}} \sum_{j'\in P^{\kappa'}} \Gamma\left(C^{\omega'}_{i'} - C^{\kappa'}_{j'} - \Delta_{\omega'\kappa'} - \delta_{r's'}\right) x^{r's'}_{\omega'i'} \left(\frac{\partial C^{\omega'}_{i'}}{\partial x^{rs}_{\omega i}} - \frac{\partial C^{\omega'}_{i'}}{\partial x^{rs}_{\omega i}}\right)$$

Combining Equations (8) and (9), we have
$$\dot{V} = \nabla V(\boldsymbol{x}) \cdot \dot{\boldsymbol{x}}$$
$$= -2\dot{\boldsymbol{x}}^T J \dot{\boldsymbol{x}} + \sum_{(r,s)\in R\times S} \sum_{\omega\in\Omega^d_{rs}} \sum_{\kappa\in\Omega^d_{rs}} \sum_{i\in P^{\omega}} \sum_{j\in P^{\kappa}} \Gamma^2\left(C^{\omega}_i - C^{\kappa}_j - \Delta_{\omega\kappa} - \delta_{rs}\right) \dot{x}^{rs}_{\omega i} \tag{10}$$

Since the link travel time of each link is continuously differentiable and non-decreasing, the marginal travel time of each link is also continuously differentiable and non-decreasing. Thereby, $J$ is positive semi-definite. We have

$$\dot{V} \leq \sum_{(r,s)\in R\times S} \sum_{\omega\in\Omega^d_{rs}} \sum_{\kappa\in\Omega^d_{rs}} \sum_{i\in P^{\omega}} \sum_{j\in P^{\kappa}} \Gamma^2\left(C^{\omega}_i - C^{\kappa}_j - \Delta_{\omega\kappa} - \delta_{rs}\right) \dot{x}^{rs}_{\omega i}$$

$$= \sum_{(r,s)\in R\times S} \sum_{\omega\in\Omega^d_{rs}} \sum_{\kappa\in\Omega^d_{rs}} \sum_{i\in P^{\omega}} \sum_{j\in P^{\kappa}} \sum_{\kappa'\in\Omega^d_{rs}} \sum_{j'\in P^{\kappa'}} \Gamma^2\left(C^{\omega}_i - C^{\kappa}_j - \Delta_{\omega\kappa} - \delta_{rs}\right) \Gamma\left(C^{\kappa'}_{j'} - C^{\omega}_i - \Delta_{\kappa'\omega} - \delta_{rs}\right) x^{rs}_{\kappa'j'} \tag{11}$$

$$- \sum_{(r,s)\in R\times S} \sum_{\omega\in\Omega^d_{rs}} \sum_{\kappa\in\Omega^d_{rs}} \sum_{i\in P^{\omega}} \sum_{j\in P^{\kappa}} \sum_{\kappa'\in\Omega^d_{rs}} \sum_{j'\in P^{\kappa'}} \Gamma^2\left(C^{\omega}_i - C^{\kappa}_j - \Delta_{\omega\kappa} - \delta_{rs}\right) \Gamma\left(C^{\omega}_i - C^{\kappa'}_{j'} - \Delta_{\omega\kappa'} - \delta_{rs}\right) x^{rs}_{\omega i}$$

Denote $\Gamma_{ij} = \Gamma\left(C^{\omega}_i - C^{\kappa}_j - \Delta_{\omega\kappa} - \delta_{rs}\right)$, for $i \in P^{\omega}, j \in P^{\kappa}, \omega, \kappa \in \Omega^d_{rs}, (r,s) \in R \times S$. Equation (11) can be rewritten as

$$\dot{V} \leq \sum_{(r,s)\in R\times S} \sum_{\omega\in\Omega^d_{rs}} \sum_{\kappa\in\Omega^d_{rs}} \sum_{i\in P^{\omega}} \sum_{j\in P^{\kappa}} \sum_{\kappa'\in\Omega^d_{rs}} \sum_{j'\in P^{\kappa'}} \Gamma^2_{ij}\left(\Gamma_{j'i} x^{rs}_{\kappa'j'} - \Gamma_{ij'} x^{rs}_{\omega i}\right)$$
$$= \sum_{(r,s)\in R\times S} \sum_{\omega\in\Omega^d_{rs}} \sum_{\kappa\in\Omega^d_{rs}} \sum_{i\in P^{\omega}} \sum_{j\in P^{\kappa}} \sum_{\kappa'\in\Omega^d_{rs}} \sum_{j'\in P^{\kappa'}} \left(\Gamma^2_{ij} - \Gamma^2_{j'j}\right) \Gamma_{j'i} x^{rs}_{\kappa'j'} \tag{12}$$

When $\Gamma_{j'i} > 0$, we have $C^{\kappa'}_{j'} - C^{\omega}_i - \Delta_{\kappa'\omega} - \delta_{rs} > 0$. Therefore,
$$\left(C^{\omega}_i - C^{\kappa}_j - \Delta_{\omega\kappa} - \delta_{rs}\right) - \left(C^{\kappa'}_{j'} - C^{\kappa}_j - \Delta_{\kappa'\kappa} - \delta_{rs}\right) = -\left(C^{\kappa'}_{j'} - C^{\omega}_i - \Delta_{\kappa'\omega}\right) < -\delta_{rs} \leq 0 \tag{13}$$

Therefore, there are three possible situations related to $\Gamma_{ij}$ and $\Gamma_{j'j}$.

$\Gamma_{ij} = 0, \Gamma_{j'j} = 0$, then $\left(\Gamma^2_{ij} - \Gamma^2_{j'j}\right) \Gamma_{j'i} x^{rs}_{\kappa'j'} = 0$;

$\Gamma_{ij} = 0, \Gamma_{j'j} > 0$, then $\left(\Gamma^2_{ij} - \Gamma^2_{j'j}\right) \Gamma_{j'i} x^{rs}_{\kappa'j'} < 0$;

$\Gamma_{ij} < 0, \Gamma_{j'j} < 0$, then
$$\left(\Gamma^2_{ij} - \Gamma^2_{j'j}\right) \Gamma_{j'i} x^{rs}_{\kappa'j'} = -\left(C^{\kappa'}_{j'} - C^{\omega}_i - \Delta_{\kappa'\omega}\right)\left(\Gamma_{ij} + \Gamma_{j'j}\right) \Gamma_{j'i} x^{rs}_{\kappa'j'} < -\delta_{rs}\left(\Gamma_{ij} + \Gamma_{j'j}\right) \Gamma_{j'i} x^{rs}_{\kappa'j'} < 0.$$

Therefore, we have
$$\frac{d}{dt} V(\boldsymbol{x}) = \nabla V(\boldsymbol{x}) \cdot \dot{\boldsymbol{x}} < 0 \tag{14}$$
except at the ALSO state, that is $\boldsymbol{x} \in E$, where
$$E = \{\boldsymbol{x} | \Gamma_{ij} x^{rs}_{\omega i} = 0, \forall i \in P^{\omega}, \forall j \in P^{\kappa}, \forall \omega, \kappa \in \Omega^d_{rs}, \forall (r,s) \in R \times S\} \tag{15}$$
□

**Theorem 4.** *The dynamical system defined in Equation (4) will always converge to the ALSO state with arbitrary initial route flows if the link travel time is continuously differentiable and non-decreasing.*

*Proof.* According to Theorems 1, 2, and Lemma 1,

$$M = \left\{ x \Big| \sum_{\omega \in \Omega_{rs}^d} \sum_{i \in P^\omega} x_{\omega i}^{rs} = D_{rs}^\tau \text{ and } x_{\omega i}^{rs} \geq 0, \forall i \in P^\omega, \forall j \in P^\kappa, \forall \omega, \kappa \in \Omega_{rs}^d, \forall (r,s) \in R \times S \right\} \quad (16)$$

is a compact set that is positively invariant with respect to the autonomous system defined in Equation (4). The candidate Lyapunov function in Equation (7) is a continuously differentiable function and $\dot{V} \leq 0$ in $M$. According to Lemma 2, the ALSO state set $E$ itself is the largest invariant set in $E$. Therefore, using LaSalle's theorem, we show that initial states with arbitrary route flows, which are always in $M$, converge to the ALSO set $E$ as $t \to \infty$. □

### 3.4. Discrete Route Assignment Dynamical System

The dynamical system described in Equation (4) is a steady state model. For implementation consideration, the dynamical system is discretized, such that the difference between the state vector in the $(k+1)^{th}$ iteration and the $k^{th}$ iteration is proportional to the local route flow changing rate in Equation (4):

$$\boldsymbol{X}_{k+1} - \boldsymbol{X}_k = \alpha_k \Phi(\boldsymbol{X}_k) \cdot \boldsymbol{X}_k \quad (17)$$

where $\alpha_k$ is the step size at the $k^{th}$ iteration. Many methods have been proposed to determine the step size in the discrete route switching day-to-day model (Powell and Sheffi, 1982; Smith and Wisten, 1995; Mounce and Carey, 2015) to guarantee the convergence property. In this study, the step size $\alpha_k$ is determined using the method proposed by Wang et al. (2019) to enhance convergence performance.

### 3.5. Numerical Illustration of Convergence Property

The Sioux Falls network is used to demonstrate the convergence performance of the proposed decentralized route flow assignment method. Suppose in the local area shown in Fig. 3(a), 4000 vehicles just pass the node 13 (on link 39) and head to the destination node 16. The local area is shown in the red dash circle, which contains three local destination nodes 23, 22, and 21. Based on the discussion of the distributed route flow assignment method, the local network can be equivalently depicted as Fig. 3(b).

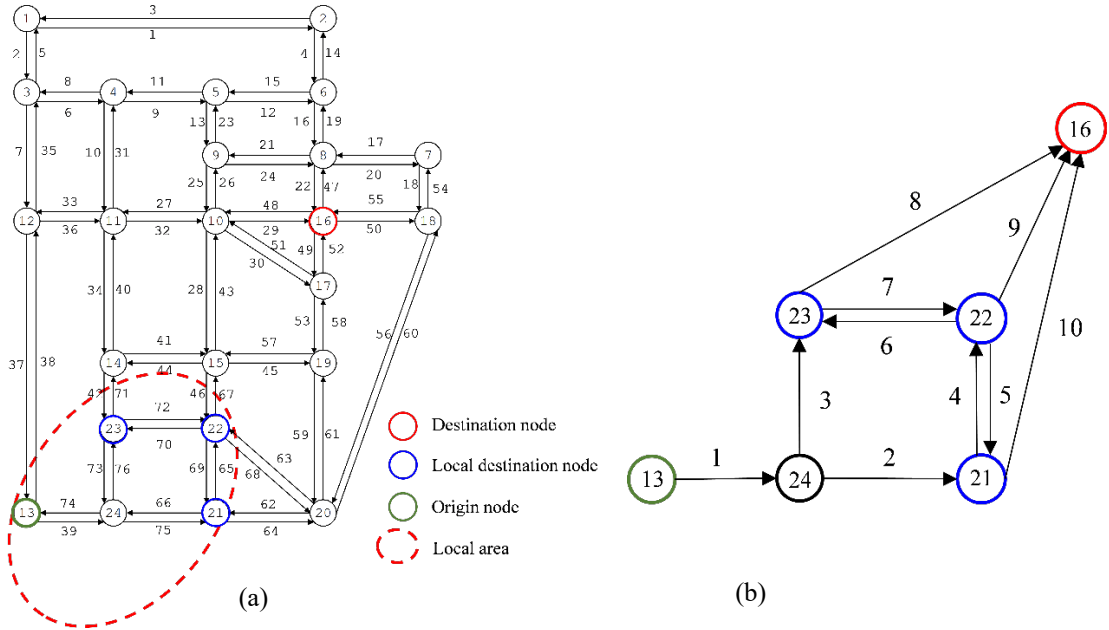

Fig. 3. Example network; (a) Local area in the Sioux Falls network; (b) Equivalent network of the local network.

The following BPR function is used to estimate the link travel time

$$t_a = t_a^0 \left[ 1 + \left(\frac{v_a}{c_a}\right)^4 \right], \forall a \quad (18)$$

where $t_a$ is the travel time of link $a$; $t_a^0$ is the free flow travel time of link $a$; and $v_a$ and $c_a$ are the flow and capacity of link $a$, respectively.

The travel time of links 8, 9, and 10 are fixed as 11, 8, and 10, respectively, since they are beyond the local network. The other inputs for the BPR function are shown in Table 1.

Table 1. Inputs of parameters in the BPR function for links in the local network.

| Links | 1 | 2 | 3 | 4 | 5 | 6 | 7 |
|---|---|---|---|---|---|---|---|
| $t_a^0$ | 3 | 6 | 5 | 2 | 2 | 3 | 3 |
| $c_a$ | 4000 | 2000 | 2000 | 2000 | 2000 | 2000 | 2000 |

To measure the solution quality, we define the convergence indicator (denoted as $G$) as follows:

$$G = \frac{\sum_{(r,s)\in R\times S}\sum_{\omega\in\Omega_{rs}^d}\sum_{\kappa\in\Omega_{rs}^d}\sum_{i\in P^\omega}\sum_{j\in P^\omega, j\neq i}\Gamma\left(C_i^\omega - C_j^\kappa - \Delta_{\omega\kappa} - \min(C_k^w, \forall k \in P^w, w \in \Omega_{rs}^d) - \delta_{rs}\right)x_{irs}^{\tau\omega}}{\sum_{(r,s)\in R\times S}\sum_{\omega\in\Omega_{rs}^d}\sum_{\kappa\in\Omega_{rs}^d}\sum_{i\in P^\omega}C_i^\omega x_{irs}^{\tau\omega}}$$

The convergence indicator implies that if the path flow solution is closer to the approximated LSO state, $G$ is closer to 0. For simplicity, assume $\Delta_{\omega\kappa} \equiv 0, \forall \omega, \kappa \in \Omega_{rs}^d$ and $\delta_{rs} = 0.1$. Fig. 4 shows the convergence performance of the proposed distributed route flow assignment method. It demonstrates that the distributed route flow assignment method can obtain a solution with the convergence criteria (i.e., the value of $G$) lower than $10^{-4}$ in 561 iterations. Thereby, this method can effectively solve the SO problem on the local network.

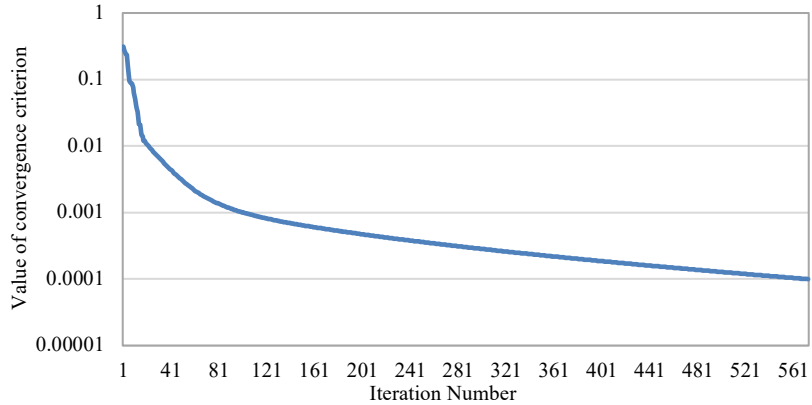

Fig. 4. Convergence performance of the proposed distributed route flow assignment method

Table 2. LSO link flows.

| Link ID | 1 | 2 | 3 | 4 | 5 | 6 | 7 | 8 | 9 | 10 |
|---|---|---|---|---|---|---|---|---|---|---|
| Flow | 4000 | 1954.4 | 2045.6 | 659.3 | 0 | 0 | 598.9 | 1446.7 | 1258.2 | 1295.1 |

## 4. Vehicle Route Assignment

The route switching dynamical system provides an optimal local route flow solution at the ALSO state set. In this section, a vehicle route assignment problem will be formulated to assign each vehicle to a specific route to achieve the optimal local route flows. To capture individual heterogeneity, the vehicle route assignment problem incorporates individual evaluation functions that define each individual's perception of the utilities of a route. The vehicle route assignment problem seeks to optimize the route assigned to each individual to maximize the sum of the utilities of all individuals. Throughout this section, we assume there exists one driver/passenger in each vehicle. Thereby, the term vehicle and individual are used interchangeably.

Suppose at time $\tau$ the initial route choices for all vehicles from origin node $r$ to destination node $s$ is denoted by $\boldsymbol{\mu} = (\mu_1, \mu_2, \ldots, \mu_n),|$ where $n$ is the demand of the global OD pair $(r,s)$, $n = |D_{rs}^\tau|$. Let $\boldsymbol{\eta} = (\eta_1, \eta_2, \ldots, \eta_n)$ be the new routes assigned to the $n$ vehicles, respectively, based on the vehicle route assignment problem. Let $v_i$ be individual $i$'s evaluation function, which measures the utility of the routes for individual $i$. For example, $v_i(\rho) = -\lambda_i T(\rho)$ is the monetary time lost for individual $i$ if he/she chooses route $\rho$, where $\lambda_i$ is the value of time of individual $i$, and $T(\rho)$ is the travel time of route $\rho$. Note that we do not have any constraints on the form of $v_i$. We only assume that the monetary loss is proportional to both the route travel time and the individual's value of time, i.e., $v_i(\rho) \sim -\lambda_i T(\rho)$. Let $\boldsymbol{v}$ be the vector of all individual evaluation functions, $\boldsymbol{v} = (v_1, v_2, \ldots, v_n)$.

To ensure each individual can accept the assigned route, let $p_i (i = 1, 2, \cdots, n)$ be the payment of individual $i$, where $p_i$ can be a negative value or a positive value, meaning that the individual will receive money (negative value) or pay some money (positive value) for the assigned route. Let $\boldsymbol{p} = (p_1, p_2, \ldots, p_n)$.

Based on the above discussion, if an individual $i$ is switching from the initial route $\mu_i$ to the new route $\eta_i$, the utilities (denoted as $u_i$) for this change is

$$u_i = v_i(\eta_i) - v_i(\mu_i) - p_i \tag{19}$$

We seek to determine the optimal new route for each individual (i.e., $\boldsymbol{\eta}^*$) such that the sum of the utilities across individuals is maximized, i.e.,

$$\boldsymbol{\eta}^* = \underset{\boldsymbol{\eta}}{\operatorname{argmax}} \sum_{i=1}^{n} u_i = \underset{\boldsymbol{\eta}}{\operatorname{argmax}} \left( \sum_{i=1}^{n} (v_i(\eta_i) - v_i(\mu_i)) - C \right) = \underset{\boldsymbol{\eta}}{\operatorname{argmax}} \sum_{i=1}^{n} v_i(\eta_i) \tag{20}$$

where $\sum_{i=1}^{n} p_i = C$ is a constant. Equation (20) indicates that the optimal route solutions for all individuals $\boldsymbol{\eta}^*$ does not depend on $p_i, \forall i$.

Let $A = \{1, 2, \ldots, n\}$ and $P_{rs}$ be the set of all individuals for global OD pair $(r, s)$ and the set of all local routes for global OD pair $(r, s)$, respectively. Denote $\rho$ as an arbitrary route in set $P_{rs}$. The vehicle route assignment problem can be formulated as the following linear integer program:

$$\max_{b_{i\rho}} \sum_{(i,\rho) \in A \times P_{rs}} b_{i\rho} v_i(\rho) \tag{21a}$$

$$\text{subject to} \sum_{i \in A} b_{i\rho} = x_\rho^{rs} \text{ for } \rho \in P_{rs}, \tag{21b}$$

$$\sum_{\rho \in P_{rs}} b_{i\rho} = 1 \text{ for } i \in A, \tag{21c}$$

$$b_{i\rho} \in \{0, 1\} \text{ for } (i, \rho) \in A \times P_{rs} \tag{21d}$$

where $b_{i\rho}$ is a binary indicator, $b_{i\rho} = 1$ if vehicle $i$ is assigned to route $\rho$. Otherwise, $b_{i\rho} = 0$. The objective of the vehicle route assignment problem (21) (Equation (21a)) seeks to maximize the sum of the utilities of the agents. Equation (21b) ensures that the vehicle route assignments can lead to the approximated optimal route flow given by the route switching dynamical system (4). Equation (21c) indicates that each vehicle can only be assigned to one route.

**Theorem 5 (Heller and Tompkins, 1956).** *Let A be an $m$ by $n$ matrix whose rows can be partitioned into two disjoint sets B and C. Then, the following four conditions together are sufficient for A to be totally unimodular:*

- *Every entry in A is $0, +1$ or $-1$;*
- *Every column of A contains at most two non-zero (i.e., $+1$ or $-1$) entries;*
- *If two non-zero entries in a column of A have the same sign, then the row of one is in B, and the other is in C;*
- *If two non-zero entries in a column of A have opposite signs, then the rows of both are in B, or both are in C.*

Let $\boldsymbol{y} = [y_1 \ y_2 \ \cdots]$ be a column vector of binary decision variables with the dimension equal to the number of columns of matrix $A$. Let $b$ be a column vector with each entry being an integer. The following theorem will be useful to demonstrate the method to obtain the solution to problem (21).

**Theorem 6 (Guzelsoy and Ralphs, 2007).** *If the constraint matrix A and the right-hand side vector b of a mixed-integer program are totally unimodular and integer respectively, then the linear integer programming problem constructed upon constraints $\{A\boldsymbol{y} = b, y_i = 0 \text{ or } 1, \forall i\}$ can be relaxed as the corresponding linear programming problem with constraints $\{A\boldsymbol{y} = b, 0 \leq y_i = 1, \forall i\}$, i.e., the optimal solution of the relaxed linear program must be integer-valued.*

It is not hard to prove that the constraint matrix of Equation (21) is unimodular. Note that the right-hand side of Equations (21b) and (21c) are all integers. Using Theorem 6, we can replace the binary constraint defined in Equation (21d) with $b_{i\rho} \in [0, 1]$ for $(i, \rho) \in A \times P_{rs}$. Thereby, the linear integer programming problem can be converted into a relaxed linear programming problem which can be solved using, for example, the simplex method.

Note that agents' payments are related to the incentives they can obtain according to the incentive mechanism in the next section. However, Equation (20)) shows that as long as the sum of incentives for the vehicle group is fixed, the optimal vehicle route assignments derived from group utility maximization does not change with the incentives received by each agent. In other words, the optimal incentives generated by the incentive mechanism does not affect the vehicle route assignment model.

The local network in Fig. 3(b) will be used to illustrate the procedure to determine the vehicle route assignments. As we want to specify the route assigned to each vehicle, the number of vehicles in the group is limited as 20. The inputs for the parameters in the link performance function are shown in Table 3. Table 4 shows the computed LSO route flows.

Table 3. Inputs of parameters in the BPR function for links in the local network.

| Links | 1 | 2 | 3 | 4 | 5 | 6 | 7 |
|---|---|---|---|---|---|---|---|
| $t_a^0$ | 3 | 6 | 5 | 2 | 2 | 3 | 3 |
| $c_a$ | 8 | 4 | 4 | 4 | 4 | 4 | 4 |

Table 4. LSO route flows.

| Route | 1 | 2 | 3 | 4 |
|---|---|---|---|---|
| Links | 1-2-10 | 1-2-4-9 | 1-3-7-9 | 1-3-8 |
| Travel time | 289.96 | 290.09 | 331.61 | 331.50 |
| Flow | 8 | 2 | 1 | 9 |

Suppose $v_i(\rho) = -\lambda_i T(\rho)$, where the value of time parameters $\lambda_i$ are randomly generated. The vehicle route assignment problem (21) is solved using the simplex method. The optimal vehicle route assignment results are shown in Table 5 in Section 5.1.

## 5. Incentive Mechanism

Similar to the route-swapping dynamical system, the vehicle route assignment problem (21) is also formulated from the system point of view, seeking to determine optimal routes for each individual to maximize the sum of utilities of the vehicle group. While the vehicle route assignment problem considers heterogeneity in individuals' preferences by incorporating individual evaluation functions, its solutions do not depend on the individual-level payments. This indicates that the travel costs of the routes assigned to the individuals are different. Thereby, the vehicle route assignments may not be acceptable to some agents as these assignments do not fully consider fairness and individual heterogeneity. Next, we will propose an incentive mechanism to promote the acceptance of the vehicle route assignments.

The incentive mechanism ensures that the vehicle route assignments are envy-free, implying each individual is satisfied with the route assigned to him/her. However, envy-freeness is achieved only if all agents are involved with the incentive mechanism voluntarily. To enable this, we provide additional group compensations to ensure each agent is expected to benefit from this incentive mechanism. Another concern is the honesty of the agents, i.e., whether agents report their information in the individual evaluation functions (e.g., the value of time) honestly. This concern will be addressed by the expectation incentive compatibility. We will show analytically that the agents will benefit the most from this incentive mechanism if they behave honestly. Through the aforementioned properties, the proposed incentive mechanism relaxes the obligation assumptions (honesty, compliance) in traveler behavior that constrain most existing studies. Individual travelers are not assumed to follow the recommended routes or behave honestly but are nudged to do so willingly.

### 5.1. Expected Envy-free Incentive Mechanism

We label vehicle route assignments as envy-free if the utility obtained by any agent through selecting another route is not higher than that of the one assigned to him/her, i.e.,

$$u_i(\eta_i, p_i) \geq u_i(\eta_j, p_i), \forall\, i, j \in A \tag{22}$$

where $\eta_i$ is the route assigned to agent $i$ by the vehicle route assignment problem; $\eta_j$ is another route (i.e., the one assigned to agent $j$).

Denote $e_{ij}$ as agent $i$'s evaluation of the new route assigned to agent $j$:

$$e_{ij} = v_i(\eta_j) - v_i(\mu_j) + a_j, \text{for } i, j \in A \tag{23}$$

where $a_j$ is the adjustment incentives given to agent $j$. Haake et al. (2002) proposed a compensation procedure which eliminates the envy of the initial utility assignment for $n = |A|$ agents. Following the procedure, we can guarantee envy-freeness in $n - 1$ compensation steps, which means we can determine $a_i, i \in A$, such that

$$e_{ii} \geq e_{ij}, \forall i,j \in A \tag{24}$$

Since we are pursuing budget balancing, the sum of incentives $\sum_i a_i$ should be paid equally by all agents, which will maintain the envy-freeness. Taking this into account, each agent pays an additional $\frac{1}{n}\sum_i a_i$, Equation (24) becomes Equation (22). In this way, agent $i$'s total payment can be derived as

$$p_i = \frac{1}{n}\sum_{j \in A} a_j - a_i \tag{25}$$

Note that $a_i$ and $p_i$ are related to the initial route choice $\mu_i$. It is reasonable to assume that in the long term, the initial state at the beginning of each time interval is random. The adjustment procedure defined above would compensate more the agents who are less satisfied with the initial routes. It is envy-free under a specific initial route choice distribution $\boldsymbol{\mu}$, but agents who have better initial routes would envy agents with worse initial routes. Since the optimal vehicle route assignments generated in the second stage are not related to the evaluation of the initial routes, agents with worse initial routes are envied more because they will have a higher compensation. Therefore, to address the unfairness related to the randomness of the initial route choice distribution, we define agent $i$'s expected evaluation of the new route assigned to agent $j$ as the expected utility of agent $i$ if he/she were assigned to agent $j$'s new route:

$$E_{\boldsymbol{\mu}}(e_{ij}) = v_i(\eta_j) - E_{\boldsymbol{\mu}}\left(v_i(\mu_j)\right) + p_j, \text{ for } i,j \in A \tag{26}$$

where $E_{\boldsymbol{\mu}}(e_{ij}) = \frac{1}{n}\sum_{k=1}^{n} v_i(\mu_k)$ is agent $i$'s expected evaluation of agent $j$'s initial routes, which is the average of agent $i$'s evaluation of all initial routes. The expected envy-freeness is represented as

$$E_{\boldsymbol{\mu}}(e_{ii}) \geq E_{\boldsymbol{\mu}}(e_{ij}), \forall i,j \in A \tag{27}$$

Based on this definition, the expected envy-free compensation procedure can be described as follows. Note that it is a computational procedure, which implies that there will not be an $n$-round monetary transfer or an extra charging step. Each agent will be notified of the incentives (can be positive or negative) associated with his/her route choice update.

Step 1. In the first round, there will always be at least one agent who experiences no envy (see Theorem 1 in Haake et al. (2002)). Therefore, no adjustment incentives are provided in the first round.
Step 2. Calculate $E_{\boldsymbol{\mu}}(e_{ij}), i,j \in A$. If all agents are expected envy-free, go to Step 4.
Step 3. Perform a new round of compensations: identify all agents who are envious of a non-envious agent the most, and give them an adjustment incentive using their maximum envy difference. Go to Step 2.
Step 4. Sum up all the adjustment incentives computed in all rounds, split them equally, and charge all agents.

Following the above steps, we compute the optimal incentives ($a_i$) and payments ($p_i$) for 20 vehicles, which are shown in Table 5 along with the route assignments $\eta_i$. Vehicle 1 is assigned to Route 1 in Table 4 (1-2-10), which is the route with the shortest travel time. However, it needs to pay its share of the sum of incentives, 12.401. Vehicle 3 is assigned to Route 4 in Table 4 (1-3-8), which has the second highest travel time. It is then compensated by 24.786 to ensure acceptance of this route. With its share of the sum of incentives also being 12.401, its total payment is 12.401-24.786 = -12.385. Also, the expected evaluation table corresponding to the vehicle route assignments and payments is shown in Table 6, where the value in row $i$ and column $j$ is agent $i$'s expected evaluation of the new route assigned to agent $j$ as defined in Equation (26). The values on the diagonal line with a shaded background in the table are the $E_{\boldsymbol{\mu}}(e_{ii})$. Note that these values are larger than the other values in the same column. Thereby, Equation (27) holds for all $i,j \in [1, \ldots, 20]$. This implies that each agent perceives the utility of the assigned route to him/her as being higher than on other routes. Therefore, the proposed incentive mechanism is expected envy-free.

Table 5. Vehicle route assignments and incentives.

| Vehicle ID | $\lambda_i$ | $\eta_i$ | $a_i$ | $p_i$ |
|---|---|---|---|---|
| 1 | 0.80 | 1 | 0.000 | 12.401 |
| 2 | 0.91 | 1 | 0.000 | 12.401 |
| 3 | 0.45 | 4 | 24.786 | -12.385 |
| 4 | 0.46 | 4 | 24.786 | -12.385 |
| 5 | 0.72 | 1 | 0.000 | 12.401 |
| 6 | 0.64 | 2 | 0.080 | 12.321 |
| 7 | 0.54 | 4 | 24.786 | -12.385 |
| 8 | 0.84 | 1 | 0.000 | 12.401 |
| 9 | 0.61 | 2 | 0.080 | 12.321 |
| 10 | 0.42 | 4 | 24.786 | -12.385 |
| 11 | 0.60 | 4 | 24.786 | -12.385 |
| 12 | 1.00 | 1 | 0.000 | 12.401 |
| 13 | 0.40 | 4 | 24.786 | -12.385 |
| 14 | 0.43 | 4 | 24.786 | -12.385 |
| 15 | 0.87 | 1 | 0.000 | 12.401 |
| 16 | 0.76 | 1 | 0.000 | 12.401 |
| 17 | 0.23 | 4 | 24.786 | -12.385 |
| 18 | 0.71 | 1 | 0.000 | 12.401 |
| 19 | 0.49 | 4 | 24.786 | -12.385 |
| 20 | 0.15 | 3 | 24.788 | -12.387 |

Table 6. Expected evaluation table

| $E_\mu(e_{ij})$ | 1 | 2 | 3 | 4 | 5 | 6 | 7 | 8 | 9 | 10 | 11 | 12 | 13 | 14 | 15 | 16 | 17 | 18 | 19 | 20 |
|---|---|---|---|---|---|---|---|---|---|---|---|---|---|---|---|---|---|---|---|---|
| 1 | 4.143 | 6.482 | -3.132 | -2.735 | 2.555 | 0.958 | -1.244 | 4.994 | 0.375 | -3.659 | 0.009 | 8.363 | -4.094 | -3.385 | 5.620 | 3.387 | -7.585 | 2.427 | -2.254 | -9.197 |
| 2 | 4.143 | 6.482 | -3.132 | -2.735 | 2.555 | 0.958 | -1.244 | 4.994 | 0.375 | -3.659 | 0.009 | 8.363 | -4.094 | -3.385 | 5.620 | 3.387 | -7.585 | 2.427 | -2.254 | -9.197 |
| 3 | -4.105 | -6.436 | 3.147 | 2.751 | -2.522 | -0.931 | 1.265 | -4.953 | -0.349 | 3.672 | 0.015 | -8.311 | 4.106 | 3.399 | -5.577 | -3.351 | 7.585 | -2.394 | 2.271 | 9.192 |
| 4 | -4.105 | -6.436 | 3.147 | 2.751 | -2.522 | -0.931 | 1.265 | -4.953 | -0.349 | 3.672 | 0.015 | -8.311 | 4.106 | 3.399 | -5.577 | -3.351 | 7.585 | -2.394 | 2.271 | 9.192 |
| 5 | 4.143 | 6.482 | -3.132 | -2.735 | 2.555 | 0.958 | -1.244 | 4.994 | 0.375 | -3.659 | 0.009 | 8.363 | -4.094 | -3.385 | 5.620 | 3.387 | -7.585 | 2.427 | -2.254 | -9.197 |
| 6 | 4.124 | 6.449 | -3.108 | -2.713 | 2.545 | 0.958 | -1.231 | 4.970 | 0.378 | -3.631 | 0.015 | 8.319 | -4.064 | -3.359 | 5.592 | 3.372 | -7.534 | 2.418 | -2.235 | -9.136 |
| 7 | -4.105 | -6.436 | 3.147 | 2.751 | -2.522 | -0.931 | 1.265 | -4.953 | -0.349 | 3.672 | 0.015 | -8.311 | 4.106 | 3.399 | -5.577 | -3.351 | 7.585 | -2.394 | 2.271 | 9.192 |
| 8 | 4.143 | 6.482 | -3.132 | -2.735 | 2.555 | 0.958 | -1.244 | 4.994 | 0.375 | -3.659 | 0.009 | 8.363 | -4.094 | -3.385 | 5.620 | 3.387 | -7.585 | 2.427 | -2.254 | -9.197 |
| 9 | 4.124 | 6.449 | -3.108 | -2.713 | 2.545 | 0.958 | -1.231 | 4.970 | 0.378 | -3.631 | 0.015 | 8.319 | -4.064 | -3.359 | 5.592 | 3.372 | -7.534 | 2.418 | -2.235 | -9.136 |
| 10 | -4.105 | -6.436 | 3.147 | 2.751 | -2.522 | -0.931 | 1.265 | -4.953 | -0.349 | 3.672 | 0.015 | -8.311 | 4.106 | 3.399 | -5.577 | -3.351 | 7.585 | -2.394 | 2.271 | 9.192 |
| 11 | -4.105 | -6.436 | 3.147 | 2.751 | -2.522 | -0.931 | 1.265 | -4.953 | -0.349 | 3.672 | 0.015 | -8.311 | 4.106 | 3.399 | -5.577 | -3.351 | 7.585 | -2.394 | 2.271 | 9.192 |
| 12 | 4.143 | 6.482 | -3.132 | -2.735 | 2.555 | 0.958 | -1.244 | 4.994 | 0.375 | -3.659 | 0.009 | 8.363 | -4.094 | -3.385 | 5.620 | 3.387 | -7.585 | 2.427 | -2.254 | -9.197 |
| 13 | -4.105 | -6.436 | 3.147 | 2.751 | -2.522 | -0.931 | 1.265 | -4.953 | -0.349 | 3.672 | 0.015 | -8.311 | 4.106 | 3.399 | -5.577 | -3.351 | 7.585 | -2.394 | 2.271 | 9.192 |
| 14 | -4.105 | -6.436 | 3.147 | 2.751 | -2.522 | -0.931 | 1.265 | -4.953 | -0.349 | 3.672 | 0.015 | -8.311 | 4.106 | 3.399 | -5.577 | -3.351 | 7.585 | -2.394 | 2.271 | 9.192 |
| 15 | 4.143 | 6.482 | -3.132 | -2.735 | 2.555 | 0.958 | -1.244 | 4.994 | 0.375 | -3.659 | 0.009 | 8.363 | -4.094 | -3.385 | 5.620 | 3.387 | -7.585 | 2.427 | -2.254 | -9.197 |
| 16 | 4.143 | 6.482 | -3.132 | -2.735 | 2.555 | 0.958 | -1.244 | 4.994 | 0.375 | -3.659 | 0.009 | 8.363 | -4.094 | -3.385 | 5.620 | 3.387 | -7.585 | 2.427 | -2.254 | -9.197 |
| 17 | -4.105 | -6.436 | 3.147 | 2.751 | -2.522 | -0.931 | 1.265 | -4.953 | -0.349 | 3.672 | 0.015 | -8.311 | 4.106 | 3.399 | -5.577 | -3.351 | 7.585 | -2.394 | 2.271 | 9.192 |
| 18 | 4.143 | 6.482 | -3.132 | -2.735 | 2.555 | 0.958 | -1.244 | 4.994 | 0.375 | -3.659 | 0.009 | 8.363 | -4.094 | -3.385 | 5.620 | 3.387 | -7.585 | 2.427 | -2.254 | -9.197 |
| 19 | -4.105 | -6.436 | 3.147 | 2.751 | -2.522 | -0.931 | 1.265 | -4.953 | -0.349 | 3.672 | 0.015 | -8.311 | 4.106 | 3.399 | -5.577 | -3.351 | 7.585 | -2.394 | 2.271 | 9.192 |
| 20 | -4.112 | -6.445 | 3.143 | 2.747 | -2.529 | -0.936 | 1.260 | -4.961 | -0.354 | 3.669 | 0.010 | -8.321 | 4.103 | 3.395 | -5.586 | -3.358 | 7.584 | -2.401 | 2.268 | 9.192 |

## 5.2. Expected Individual Rationality with Budget Balance

The individual rationality ensures that CAVs want to participate as they are at least as well off by participating compared with if they do not participate:

$$u_i = v_i(\eta_i) - v_i(\mu_i) - p_i \geq 0, \text{for } i \in A \tag{28}$$

Note that the incentive mechanism may not exist to increase the utilities of all agents as someone may be assigned a very good route initially. To address this problem, we introduce the definition of expected individual rationality. The incentive mechanism is expected individual rational if it can increase the expected utilities for all agents, i.e.,

$$E_\mu(u_i) = v_i(\eta_i) - E_\mu(v_i(\mu_i)) - p_i \geq 0, \text{for } i \in A \tag{29}$$

**Theorem 7.** *Suppose the expected evaluations of the new routes of all agents are higher than those of their old ones, i.e., $E_\eta(v_i) \geq E_\mu(v_i), i \in A$, then the incentive mechanism defined above is expected individual rational.*

*Proof.* From Equation (27), we have:
$$v_i(\eta_i) + a_i \geq v_i(\eta_j) + a_j, \forall i, j \in A.$$
Summing over $j, j \in A$, we have
$$n(v_i(\eta_i) + a_i) \geq \sum_{j \in A}(v_i(\eta_j) + a_j)$$
$$v_i(\eta_i) + a_i - \frac{1}{n}\sum_{j \in A} a_j \geq E_\eta(v_i)$$
$$u_i + v_i(\mu_i) \geq E_\eta(v_i)$$
Therefore,
$$E_\mu(u_i) \geq E_\eta(v_i) - E_\mu(v_i) \geq 0 \qquad \square$$

Theorem 7 implies that vehicles are willing to participate in the mechanism if they perceive the new route options developed using the route swapping dynamical system are better, on average. However, the route swapping dynamical system cannot reduce the travel times of vehicles for each global OD pair as it seeks to reach the ALSO state. In other words, the travel times of some vehicles will increase if they follow the routes provided by the vehicle route assignment problem (21). Thereby, they may not be willing to participate. To address this problem, we now add an additional group compensation process to charge the vehicles that benefit from the route switching, and to compensate those whose travel times are increased after route switching. The objective of the compensation is to achieve the expected individual rationality for everyone.

Consider the situation where $v_i(\rho) = -\lambda_i T(\rho)$. The condition for expected individual rationality in Theorem 7 can be written as
$$\lambda_i(\bar{T}_\mu - \bar{T}_\eta) \geq 0 \tag{30}$$
where $\bar{T}_\mu, \bar{T}_\eta$ are the average travel times of the initial route options $\boldsymbol{\mu}$ and new route options $\boldsymbol{\eta}$. For each agent in the vehicle group, we provide an additional group compensation, $\lambda_i(\bar{T}_\eta - \bar{T}_\mu + \epsilon)$, where $\epsilon \geq 0$. Then, agent $i$'s expected evaluation of agent $j$'s new route becomes
$$E_\mu(e_{ij}) = -\lambda_i T(\eta_i) + \lambda_i \bar{T}_\mu + a_j + \lambda_i(\bar{T}_\eta - \bar{T}_\mu + \epsilon) = -\lambda_i T'(\eta_i) + \lambda_i \bar{T}_\mu + a_j \tag{31}$$
where $T'(\eta_i) = T(\eta_i) + \bar{T}_\mu - \bar{T}_\eta - \epsilon$ is the adjusted travel time of $\eta_i$. Replacing $\bar{T}_\eta$ with $\bar{T}'_\eta = \frac{1}{n}\sum_{i=1}^n T(\eta_i)$ in Equation (30), the additional group compensation ensures the expected individual rationality condition:
$$\lambda_i(\bar{T}_\mu - \bar{T}'_\eta) = \lambda_i \epsilon \geq 0 \tag{32}$$
The sum of all payments by agents in Equation (20) in the vehicle route assignment model can be formulated as:
$$\sum_{i=1}^n p_i = -\left(\sum_{i=1}^n \lambda_i\right)(\bar{T}_\eta - \bar{T}_\mu + \epsilon) = -n\bar{\lambda}(\bar{T}_\eta - \bar{T}_\mu + \epsilon) \tag{33}$$
Note that the sum of all payments is a constant; thus, the optimal vehicle route assignments do not change after we introduce the additional group compensations. When $\epsilon = 0$, $\sum_{i=1}^n p_i$ is positive if $\bar{T}_\eta < \bar{T}_\mu$, and it is negative if $\bar{T}_\eta > \bar{T}_\mu$. This indicates that the traffic manager collects money from the vehicle groups whose average travel time decreases and pays the vehicle groups whose average travel time increases. Let us assume that the average value of time $\bar{\lambda}$ is the same for all vehicle groups. Then, the traffic manager can achieve the expected individual rationality and can make a profit in the long term if the total travel time of all vehicle groups decreases (same expected individual utility for agents participating and agents not participating when $\epsilon = 0$). We can adjust $\epsilon$ to be a positive value to realize budget balance and ensure that each agent participating in the proposed incentive mechanism can benefit (i.e., the utility for switching to the assigned route is positive).

*5.3. Expected Incentive Compatibility*

In economics, incentive compatibility constraint motivates agents to behave in a manner consistent with the optimal solution. In our incentive mechanism, the objective is that the utilities agents can obtain by reporting their true route evaluation functions (i.e., $v_i, \forall i$) are larger than by reporting arbitrary route evaluation functions (denoted as $v'_i$), to promote honest behavior. That is, the incentive compatibility constraint seeks to ensure
$$u_i\big((v_i, \boldsymbol{v} - v_i)\big) \geq u_i\big((v'_i, \boldsymbol{v} - v_i)\big), \forall i \tag{34}$$
where $\boldsymbol{v} - v_i$ denotes the route evaluation function combination of other vehicles except for agent $i$. However, Green and Laffont (1979) have shown that non-manipulability is incompatible with envy-freeness and the budget balance constraint. Andersson et al. (2014) compromise on the non-manipulability by seeking the least-manipulable mechanisms among the envy-free and budget-balanced ones. They define a new measure of minimal manipulability as the number of agents who can manipulate the rule at a given preference profile and show that we could obtain optimal fair allocation rules as agents-counting-minimally manipulable rules. Another approach is to weaken or abandon the budget balance constraint. Cohen et al. (2010) prove that we could determine the payments to satisfy the envy-freeness and incentive compatibility constraints separately by finding the shortest paths in a weighted directed graph. Moreover, they show that by removing some of the edges in the graph, the shortest paths can represent the payments that are both envy-free and incentive compatible. Sun and Yang (2003) replace the budget-balance constraint with a maximum payment limit for each indivisible object. They develop an allocation mechanism that fairly assigns the objects and always elicits honest preferences over both objects and money (incentive compatibility).

In our study, we modify the incentive compatibility defined in Equation (34) into the expected incentive compatibility, which indicates that agents will have no better expected utilities by manipulating their preferences. Suppose $v_i(\rho) = -\lambda_i T(\rho), \lambda_i \in [\lambda_l, \lambda_u]$, where $\lambda_i > 0$ is the time value of $i$, $T(\rho)$ is the travel time of path $\rho$, and $\lambda_l, \lambda_u$ are the lower-bound and upper-bound of $\lambda_i$, respectively. Next, we show that agents cannot benefit by reporting a manipulated $\lambda_i$ under the proposed expected envy-free compensation mechanism.

**Lemma 3.** *If $v_i(\rho) = -\lambda_i T(\rho), \forall i$, then the optimal sum of utilities defined by Equation (20) is*

$$\max_{\eta} \sum_{i=1}^{n} u_i = -\sum_{r=1}^{n} \lambda^{(r)} T^{(r)} + F(\boldsymbol{v}, \boldsymbol{\mu}) \tag{35}$$

where $\lambda^{(r)}$ is the $r^{th}$ largest $\lambda$ among all agents, $T^{(r)}$ is the $r^{th}$ shortest travel time among all agent routes and $F(\boldsymbol{v}, \boldsymbol{\mu})$ is a constant denoting the sum of the evaluations of agents' initial routes.

*Proof.* Note that $\sum_{i=1}^{n} a_i b_i$ is maximum when both $\{a_i\}$ and $\{b_i\}$ are sorted in ascending order or in descending order. Therefore, the maximum sum of utilities is achieved when the $r^{th}$ largest $\lambda$ is paired with the $r^{th}$ shortest travel time. $\square$

**Lemma 4.** *If the optimal vehicle route assignment problem assigns the agent with the value of time $\lambda^{(r)}$ to the route with travel time $T^{(r)}$, then the payment of the agent can be formulated as:*

$$p^{(1)} = \frac{1}{n} \sum_{j=2}^{n} \sum_{m=2}^{j} \lambda^{(m)} (T^{(m)} - T^{(m-1)}),$$
$$p^{(i)} = \frac{1}{n} \sum_{j=2}^{n} \sum_{m=2}^{j} \lambda^{(m)} (T^{(m)} - T^{(m-1)}) - \sum_{m=2}^{i} \lambda^{(m)} (T^{(m)} - T^{(m-1)}) \tag{36}$$

*Proof.* For notational simplicity, we re-number the agents from the largest $\lambda$ to the minimum. We denote the adjustment incentive for the $j^{th}$ agent in the $k^{th}$ round as $a_j^{[k]}$. In the first round, $a_j^{[1]} = 0, j \in A$, we have

$$E_{\boldsymbol{\mu}}(e_{11}) = \lambda^{(1)}(\bar{T}(\boldsymbol{\mu}) - T^{(1)}) + a_1^{[1]} \geq \lambda^{(1)}(\bar{T}(\boldsymbol{\mu}) - T^{(j)}) + a_j^{[1]} = E_{\boldsymbol{\mu}}(e_{1j}), j \in A$$

Therefore, agent 1 is expected envy-free. In round 2, agent $j(j > 1)$ is envious of agent 1 the most. Therefore, $a_j^{[2]} = \lambda^{(j)}(T^{(j)} - T^{(1)}), j \geq 2$, we have

$$E_{\boldsymbol{\mu}}(e_{22}) = \lambda^{(2)}(\bar{T}(\boldsymbol{\mu}) - T^{(2)}) + a_2^{[2]} \geq \lambda^{(2)}(\bar{T}(\boldsymbol{\mu}) - T^{(1)}) = E_{\boldsymbol{\mu}}(e_{21});$$
$$E_{\boldsymbol{\mu}}(e_{22}) = \lambda^{(2)}(\bar{T}(\boldsymbol{\mu}) - T^{(2)}) + a_2^{[2]} \geq \lambda^{(2)}(\bar{T}(\boldsymbol{\mu}) - T^{(j)}) + a_j^{[2]} = E_{\boldsymbol{\mu}}(e_{2j}), j \geq 2.$$

Agent 2 becomes expected envy-free. While

$$E_{\boldsymbol{\mu}}(e_{11}) = \lambda^{(1)}(\bar{T}(\boldsymbol{\mu}) - T^{(1)}) + a_1^{[1]} \geq \lambda^{(1)}(\bar{T}(\boldsymbol{\mu}) - T^{(j)}) + \lambda^{(j)}(T^{(j)} - T^{(1)}) = E_{\boldsymbol{\mu}}(e_{1j}), j \in A$$

which indicates agent 1 remains expected envy-free.

Similarly, in Round $k(k > 2)$, agent $j(j \geq k)$ is envious of agent $k-1$ the most. Therefore, $a_j^{[k]} = a_{k-1}^{[k-1]} + \lambda^{(j)}(T^{(j)} - T^{(k-1)}) = \sum_{m=2}^{k-1} \lambda^{(m)}(T^{(m)} - T^{(m-1)}) + \lambda^{(j)}(T^{(j)} - T^{(k-1)}), j \geq k$, we have

$$E_{\boldsymbol{\mu}}(e_{kk}) = \lambda^{(k)}(\bar{T}(\boldsymbol{\mu}) - T^{(k)}) + a_k^{[k]}$$
$$= \lambda^{(k)}(\bar{T}(\boldsymbol{\mu}) - T^{(k)}) + \sum_{m=j+1}^{k} \lambda^{(m)}(T^{(m)} - T^{(m-1)}) + a_j^{[j]}$$
$$\geq \lambda^{(k)}(\bar{T}(\boldsymbol{\mu}) - T^{(k)}) + \lambda^{(k)} \sum_{m=j+1}^{k} (T^{(m)} - T^{(m-1)}) + a_j^{[j]}$$
$$= \lambda^{(k)}(\bar{T}(\boldsymbol{\mu}) - T^{(j)}) + a_j^{[j]}$$
$$= E_{\boldsymbol{\mu}}(e_{kj}), \forall j < k;$$
$$E_{\boldsymbol{\mu}}(e_{kk}) = \lambda^{(k)}(\bar{T}(\boldsymbol{\mu}) - T^{(k)}) + a_k^{[k]} \geq \lambda^{(k)}(\bar{T}(\boldsymbol{\mu}) - T^{(j)}) + a_j^{[k]} = E_{\boldsymbol{\mu}}(e_{kj}), \forall j \geq k.$$

Agent $k$ becomes expected envy-free. While, for $j < k, s \geq k$,

$$E_{\boldsymbol{\mu}}(e_{jj}) = \lambda^{(j)}(\bar{T}(\boldsymbol{\mu}) - T^{(j)}) + a_j^{[j]}$$
$$= \lambda^{(j)}(\bar{T}(\boldsymbol{\mu}) - T^{(s)}) + a_j^{[j]} + \lambda^{(j)} \sum_{m=j+1}^{s} (T^{(m)} - T^{(m-1)})$$
$$= \lambda^{(j)}(\bar{T}(\boldsymbol{\mu}) - T^{(s)}) + a_j^{[j]} + \lambda^{(j)} \sum_{m=j+1}^{k-1} (T^{(m)} - T^{(m-1)}) + \lambda^{(j)}(T^{(s)} - T^{(k-1)})$$

$$\geq \lambda^{(j)}\big(\bar{T}(\boldsymbol{\mu}) - T^{(s)}\big) + a_j^{[j]} + \lambda^{(j)} \sum_{m=j+1}^{k-1} \big(T^{(m)} - T^{(m-1)}\big) + \lambda^{(s)}\big(T^{(s)} - T^{(k-1)}\big)$$

$$= \lambda^{(j)}\big(\bar{T}(\boldsymbol{\mu}) - T^{(s)}\big) + a_s^{[k]}$$

$$= E_{\boldsymbol{\mu}}(e_{js})$$

which indicates agent $j (j < k)$ stays expected envy-free. Therefore, after $k$ rounds, agents $1, 2, \ldots, k$ become expected envy-free. At the end of round $n$, we have final adjustment incentives $a_i$ which make all agents expected envy-free:

$$\begin{aligned} a_1 &= a_1^{[1]} = 0; \\ a_i &= a_i^{[i]} = \sum_{m=2}^{i} \lambda^{(m)}\big(T^{(m)} - T^{(m-1)}\big), i \geq 2. \end{aligned} \tag{37}$$

Then, we can derive the payments from $p_i = \frac{1}{n}\sum_{j=1}^{n} a_j - a_i, i \in A$:

$$p_1 = \frac{1}{n} \sum_{j=2}^{n} \sum_{m=2}^{j} \lambda^{(m)}\big(T^{(m)} - T^{(m-1)}\big),$$

$$p_i = \frac{1}{n} \sum_{j=2}^{n} \sum_{m=2}^{j} \lambda^{(m)}\big(T^{(m)} - T^{(m-1)}\big) - \sum_{m=2}^{i} \lambda^{(m)}\big(T^{(m)} - T^{(m-1)}\big).$$

□

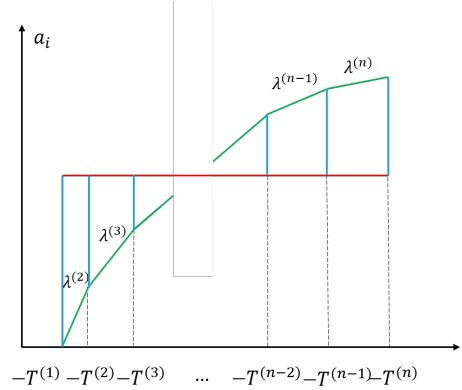

Fig. 5. Relationship between $a_i$ and $p_i$.

The result of Lemma 4 is also illustrated in Fig. 5. The green line represents the adjustment incentives that eliminate the expected envy, while the red line shows the equal share that each agent needs to pay for the incentives. Combining them, the blue lines in the figure represent the total payment each agent makes (blue lines below the red line are positive payments, while those above are negative). Now, we can analyze the additional utility an agent can gain through manipulation.

**Lemma 5. (Proof in Appendix A).** *Let $T^{(0)} = T^{(1)}$ for simplicity. If the agent with the $r^{th}$ largest $\lambda$ reports a fake value of time $\lambda_f$ which ranks $k^{th}$ among all $\lambda$s, the additional utility $(\Delta u_{(k)}^{(r)}(\lambda_f))$ he gains is*

If $r > k$:

$$\begin{aligned} \Delta u_{(k)}^{(r)}(\lambda_f) &= \frac{k-1}{n}\big(\lambda_f - \lambda^{(k)}\big)\big(T^{(k)} - T^{(k-1)}\big) \\ &\quad - \sum_{j=k}^{r-1} \big(\lambda^{(j)} - \lambda^{(r)}\big)\big(T^{(j+1)} - T^{(j)}\big) - \sum_{j=k}^{r-1} \frac{n-j}{n}\big(\lambda^{(j)} - \lambda^{(j+1)}\big)\big(T^{(j+1)} - T^{(j)}\big); \end{aligned} \tag{38a}$$

if $r = k$:

$$\Delta u_{(k)}^{(r)}(\lambda_f) = \frac{r-1}{n}\big(\lambda_f - \lambda^{(r)}\big)\big(T^{(r)} - T^{(r-1)}\big); \tag{38b}$$

if $r < k$:

$$\Delta u_{(k)}^{(r)}(\lambda_f) = -\left(\lambda^{(r)} - \frac{k-1}{n}\lambda_f - \frac{n+1-k}{n}\lambda^{(k)}\right)\left(T^{(k)} - T^{(k-1)}\right)$$
$$-\sum_{j=r}^{k-1}(\lambda^{(r)} - \lambda^{(j+1)})(T^{(j)} - T^{(j-1)}) + \sum_{j=r}^{k-1}\frac{n-j+1}{n}(\lambda^{(j)} - \lambda^{(j+1)})(T^{(j)} - T^{(j-1)}). \quad (38c)$$

From Equation (38), we note that $\Delta u_{(k)}^{(r)}(\lambda_f)$ is also related to the distribution of $\lambda^{(j)}(j \neq r)$, and $T^{(j)}(j \in A)$, which are unknown to the agent with the $r^{th}$ largest $\lambda$. Therefore, different from the definition of expected envy-freeness and expected individual rationality, the definition of expected incentive compatibility is not only related to the distribution of $\mu$ but also the distribution of $\eta$ and $\lambda = (\lambda_1, \lambda_2, \ldots, \lambda_n)$. Denote $\lambda_g$ as the real value of time for the agent. Note that in Equation (38), $\lambda^{(r)}$ is actually $\lambda_g$, which is also a variable when defining expected incentive compatibility. Therefore, the expected incentive compatibility can be defined as:

$$E_{\lambda,\mu,\eta}\left(\Delta u_{(k)}^{(r)}(\lambda_f, \lambda_g)\right) \leq 0 \quad (39)$$

**Theorem 8. (Proof in Appendix A).** *Suppose $\lambda_i \sim U(\lambda_l, \lambda_u), i \in A$, where $\lambda_l, \lambda_u$ are the lower-bound and upper-bound of $\lambda_i$, respectively. If an agent with $\lambda_g$ reports a fake value of time $\lambda_f$, the additional expected utility he/she will gain is non-positive if $\lambda_f - \lambda_g \leq 0$ or $\lambda_f - \lambda_g \geq \frac{2}{n}(\lambda_u - \lambda_l)$, where n is the number of agents.*

Theorem 8 implies that under the above assumptions, agents can only have a chance to gain positive expected additional utility by reporting a $\lambda_f$ greater than his real value of time $\lambda_g$, but less than $\lambda_g + \frac{2}{n}(\lambda_u - \lambda_l)$. It is reasonable to assume that in practice, there is a minimum division value in the value of time settings (akin to a minimum division value equal to \$0.01 when making a transfer). If the minimum division value is greater than or equal to $\frac{2}{n}(\lambda_u - \lambda_l)$, no value of time manipulation would be beneficial for all agents.

**Corollary 1.** *If agents have to report their value of time from the feasible value of time set, $\left\{\frac{j\lambda_l+(N-j)\lambda_u}{N}\Big| j = 0, 1, \ldots, N\right\}$, when N is large, the proposed incentive mechanism is expected incentive compatible for $n \geq 2N$, which implies that agents will always report the one larger than but closest to their true value of time.*

*Proof.* The minimum division value of the above feasible value of time set is $\frac{\lambda_u - \lambda_l}{N}$. When $n \geq 2N$, the minimum division value is greater than or equal to $\frac{2}{n}(\lambda_u - \lambda_l)$. For a large $N$, the reported value of time $\lambda_i$ still approximately follows the uniform distribution; thus, Theorem 8 holds. Therefore, the proposed mechanism is expected incentive compatible with this condition. □

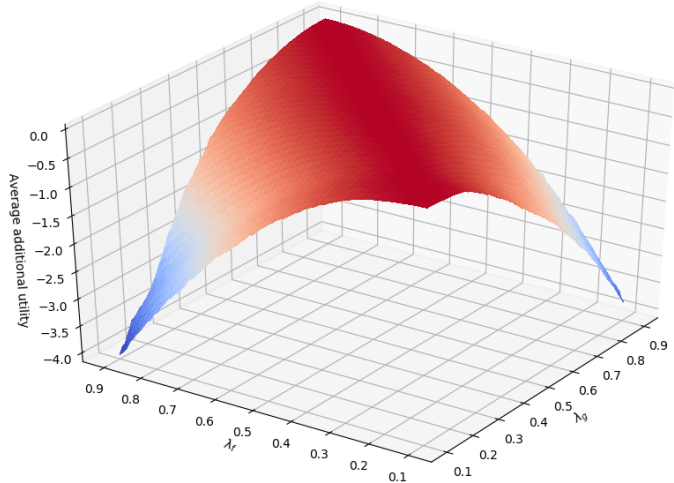

Fig. 6. Average additional utility that agent with $\lambda_g$ gains by reporting $\lambda_f$.

In practice, the number of vehicles in the vehicle group with the same global OD pair is large enough to satisfy the condition in Corollary 1 in most cases. We now illustrate the expected incentive compatibility with a numerical experiment by calculating the average additional utility that agents can gain by reporting different $\lambda_f$. Since we seek to illustrate that when $\lambda_f = \lambda_g$, the average additional utility achieves the maximum value 0, the unit does not matter in the verification. We assume that agents' values of time follow the uniform distribution $U(0.1, 0.9)$, and the minimum division value is 0.01. According to Corollary 1, the number of vehicles in the vehicle group is set as $n = 2 \times \frac{0.9-0.1}{0.01} = 160$. We randomly generate 160 route travel times from $U(50, 60)$. For each $(\lambda_g, \lambda_f)$ pair, we generate 159 other $\lambda$s randomly from $U(0.1, 0.9)$, follow the vehicle route assignment model and the incentive mechanism to calculate the utility of the agent when he/she reports $\lambda_g$ honestly and the utility when he reports $\lambda_f$, and compute additional utility gains. We repeat the procedure 100 times to calculate the average additional utility for each $(\lambda_g, \lambda_f)$ pair and plot the result in Fig. 6. It shows that agents cannot gain positive average additional utility by reporting manipulated $\lambda_f$; when $\lambda_f = \lambda_g$ (the agent is behaving honestly), the average additional utility reaches the maximum, 0.

## 6. Concluding Comments

This study proposes an incentive-based decentralized routing strategy for CAVs using information propagation in a local area. The dynamic traffic network is decomposed into deterministic network problems in small time intervals. In each time interval, following a decentralized three-stage scheme, vehicles update their route choice and take the corresponding incentives. In the first stage, a decentralized route assignment model based on a local route switching dynamical system is developed to obtain an optimal route flow solution for the ALSO state. Then, a vehicle route assignment problem is formulated to assign each vehicle a route to achieve the ALSO state and to maximize the sum of the utilities based on individual evaluation functions. Then, we propose an expected envy-free incentive mechanism to charge or compensate the agents so that everyone can accept the assigned route determined by the optimal vehicle route assignment problem. We further analyze the expected individual rationality, budget balance, and expected incentive compatibility of the incentive mechanism. The budget balance constraint is used in this study to highlight incentive strategies that do not require external funding. Instead, they entail the exchange of payments involving CAV-based individual travelers (agents) whose behavioral preferences are input into their vehicles *a priori*. Thereby, payments are seamlessly exchanged by the CAVs through V2V and V2I communications in real-time.

To the best of our knowledge, this is the first attempt to bridge the gap between heterogeneous individual-level objectives and system-level objectives in the context of vehicular routing decisions for OD travel by ensuring practical realism. The proposed decentralized routing strategy can enhance system performance and ensure individual satisfaction simultaneously by incorporating a route assignment model with an incentive mechanism. Application of the routing strategy requires the availability of local real-time traffic information, which can be realized in connected and autonomous driving environments. Moreover, the proposed routing strategy can be solved analytically and implemented in a fully decentralized manner to circumvent the computational issue that limits most existing approaches in practice. Further, we theoretically prove that the proposed incentive mechanism satisfies the expected individual rationality constraint, the budget-balance constraint, and the expected incentive compatibility constraint simultaneously, which enhances its practical applicability and realism. In summary, the proposed decentralized routing strategy is deployable both in terms of computational tractability and behavioral realism.

The study opens a new venue related to incentive/pricing strategies that leverage emerging connectivity and automation technologies. There are opportunities for future enhancements related to this study, including: (i) integrating the ALSO-based routing strategies and the incentive mechanism to analyze system performance; (ii) proposing a modeling framework to optimize the values of cover range $d$ and time interval $\tau$ to improve system performance; (iii) extending the context to a multimodal traffic environment; (iv) analyzing the expected individual rationality and expected incentive compatibility for general individual evaluation functions instead of the specific form $v_i(\rho) = -\lambda_i T(\rho)$ used in this study; (v) using cell/link transmission model instead of the BPR function to better characterize the travel time in both uncongested and congested traffic flow environments; and (vi) incorporating a reputation system to further strengthen the incentive compatibility (to ensure that agents do take the routes determined for them rather than just report their preferences honestly).

## Acknowledgements

This study is supported by funding from the National Science Foundation (1662692-CMMI) and Georgia Institute of Technology to the second author. Additional support is provided to the third author from the Natural Science Foundation of China (52002191) and Natural Science Foundation of Zhejiang province (LQ20E08004). Any errors or omissions remain the sole responsibility of the authors.

## Appendix A. Proofs of Lemma 5 and Theorem 8

**Lemma 5.** *Let $T^{(0)} = T^{(1)}$ for simplicity. If the agent with the $r^{th}$ largest $\lambda$ reports a fake value of time $\lambda_f$ which ranks $k^{th}$ among all $\lambda$s, the additional utility ($\Delta u_{(k)}^{(r)}(\lambda_f)$) he gains is*

If $r > k$:
$$\Delta u_{(k)}^{(r)}(\lambda_f) = \frac{k-1}{n}(\lambda_f - \lambda^{(k)})(T^{(k)} - T^{(k-1)})$$
$$- \sum_{j=k}^{r-1}(\lambda^{(j)} - \lambda^{(r)})(T^{(j+1)} - T^{(j)}) - \sum_{j=k}^{r-1}\frac{n-j}{n}(\lambda^{(j)} - \lambda^{(j+1)})(T^{(j+1)} - T^{(j)}); \tag{38a}$$

if $r = k$:
$$\Delta u_{(k)}^{(r)}(\lambda_f) = \frac{r-1}{n}(\lambda_f - \lambda^{(r)})(T^{(r)} - T^{(r-1)}); \tag{38b}$$

if $r < k$:
$$\Delta u_{(k)}^{(r)}(\lambda_f) = -\left(\lambda^{(r)} - \frac{k-1}{n}\lambda_f - \frac{n+1-k}{n}\lambda^{(k)}\right)(T^{(k)} - T^{(k-1)})$$
$$- \sum_{j=r}^{k-1}(\lambda^{(r)} - \lambda^{(j+1)})(T^{(j)} - T^{(j-1)}) + \sum_{j=r}^{k-1}\frac{n-j+1}{n}(\lambda^{(j)} - \lambda^{(j+1)})(T^{(j)} - T^{(j-1)}). \tag{38c}$$

*Proof.* The utility difference consists of three parts: the evaluation difference between $T^{(r)}$ and $T^{(k)}$, the adjustment incentive difference for $r^{th}$ largest and $k^{th}$ largest $\lambda$, and the difference in $\frac{1}{n}\sum_{j=1}^{n} a_j$ under $(\lambda^{(1)}, \lambda^{(2)}, \ldots, \lambda^{(r)}, \ldots, \lambda^{(n)})$ and $(\lambda^{(1)}, \ldots, \lambda_f, \ldots, \lambda^{(n)})$.

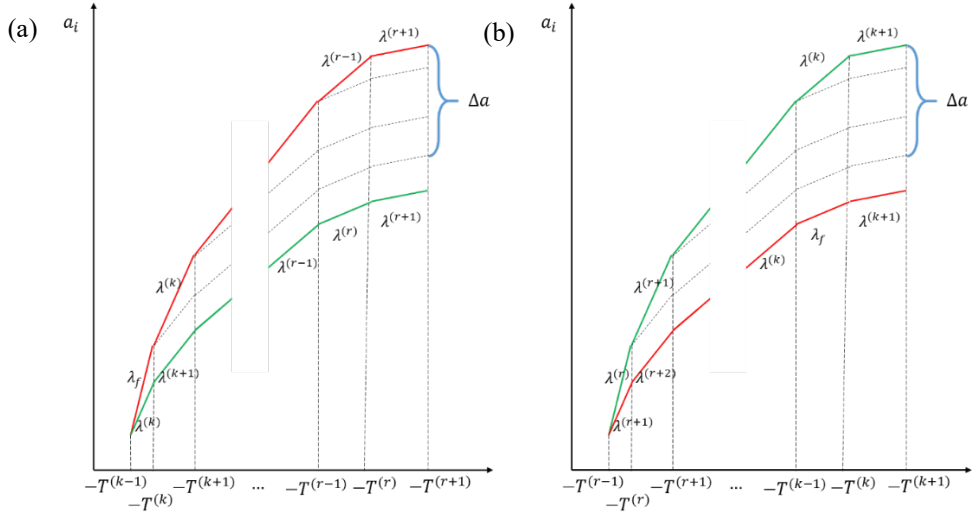

Fig. 7. Adjustment incentive difference for (a) $r > k$; (b) $r < k$.

For $r > k$,
$$\Delta u_{(k)}^{(r)}(\lambda_f) = [\lambda^{(r)}(T^{(r)} - T^{(k)})] + \left[\lambda_f(T^{(k)} - T^{(k-1)}) - \sum_{j=k}^{r}\lambda^{(j)}(T^{(j)} - T^{(j-1)})\right]$$
$$- \frac{1}{n}\left[(\lambda_f - \lambda^{(k)})(T^{(k)} - T^{(k-1)})(n+1-k) + \sum_{j=k}^{r-1}(\lambda^{(j)} - \lambda^{(j+1)})(T^{(j+1)} - T^{(j)})(n-j)\right]$$
$$= \frac{k-1}{n}(\lambda_f - \lambda^{(k)})(T^{(k)} - T^{(k-1)})$$

$$-\sum_{j=k}^{r-1}(\lambda^{(j)}-\lambda^{(r)})(T^{(j+1)}-T^{(j)})-\sum_{j=k}^{r-1}\frac{n-j}{n}(\lambda^{(j)}-\lambda^{(j+1)})(T^{(j+1)}-T^{(j)});$$

For $r = k$,

$$\Delta u_{(k)}^{(r)}(\lambda_f) = 0 + [(\lambda_f - \lambda^{(r)})(T^{(r)} - T^{(r-1)})] - \frac{1}{n}[(\lambda_f - \lambda^{(r)})(T^{(r)} - T^{(r-1)})(n+1-r)]$$

$$= \frac{r-1}{n}(\lambda_f - \lambda^{(r)})(T^{(r)} - T^{(r-1)});$$

For $r < k$,

$$\Delta u_{(k)}^{(r)}(\lambda_f) = [\lambda^{(r)}(T^{(r)} - T^{(k)})] + \left[\lambda_f(T^{(k)} - T^{(k-1)}) + \sum_{j=r}^{k-1}\lambda^{(j+1)}(T^{(j)} - T^{(j-1)}) - \lambda^{(r)}(T^{(r)} - T^{(r-1)})\right]$$

$$+ \frac{1}{n}\left[(\lambda^{(k)} - \lambda_f)(T^{(k)} - T^{(k-1)})(n+1-k) + \sum_{j=r}^{k-1}(\lambda^{(j)} - \lambda^{(j+1)})(T^{(j)} - T^{(j-1)})(n-j+1)\right]$$

$$= -\left(\lambda^{(r)} - \frac{k-1}{n}\lambda_f - \frac{n+1-k}{n}\lambda^{(k)}\right)(T^{(k)} - T^{(k-1)})$$

$$- \sum_{j=r}^{k-1}(\lambda^{(r)} - \lambda^{(j+1)})(T^{(j)} - T^{(j-1)}) + \sum_{j=r}^{k-1}\frac{n-j+1}{n}(\lambda^{(j)} - \lambda^{(j+1)})(T^{(j)} - T^{(j-1)}).$$

□

**Theorem 8.** *Suppose $\lambda_i \sim U(\lambda_l, \lambda_u), i \in A$, where $\lambda_l, \lambda_u$ are the lower-bound and upper-bound of $\lambda_i$, respectively. If an agent with $\lambda_g$ reports a fake value of time $\lambda_f$, the additional expected utility he/she will gain is non-positive if $\lambda_f - \lambda_g \leq 0$ or $\lambda_f - \lambda_g \geq \frac{2}{n}(\lambda_u - \lambda_l)$, where $n$ is the number of agents.*

*Proof.* When $\lambda_f = \lambda_g = \lambda^{(r)}$, $\Delta u_{(k)}^{(r)}(\lambda_f, \lambda_g) = 0$, we have

$$E_{\lambda,\mu,\eta}\left(\Delta u_{(k)}^{(r)}(\lambda_f, \lambda_g)\right) = 0. \tag{40}$$

When $\lambda_f < \lambda_g = \lambda^{(r)}$, the agent does not know the distribution of $\lambda$ and $\mu$, and hence does not know how large $\lambda_f$ and $\lambda_g$ are.

$$E_{\lambda,\mu,\eta}\left(\Delta u_{(k)}^{(r)}(\lambda_f, \lambda_g)\right) = \int_\lambda \int_\eta \sum_{r=1}^n \sum_{k=r}^n P_{r,k}^{n-1} P_{\lambda|r,k}^{n-1} P_\eta \Delta u_{(k)}^{(r)}(\lambda_f, \lambda_g) d\lambda d\eta \tag{41}$$

where $P_{r,k}^{n-1} = P(\lambda_f \in (\lambda^{(k-1)}, \lambda^{(k)}], \lambda_g \in (\lambda^{(r-1)}, \lambda^{(r+1)}])$ is the probability that $\lambda_f$ is larger than or equal to $\lambda^{(k)}$ but smaller than $\lambda^{(k-1)}$ while $\lambda_g$ is larger than or equal to $\lambda^{\wedge}((r+1))$ but smaller than $\lambda^{(r-1)}$ when there are $n-1$ other agents, $P_{\lambda|r,k}^{n-1}$ is the probability distribution of $\lambda$ conditioned on $\lambda_f \in (\lambda^{(k-1)}, \lambda^{(k)}]$ and $\lambda_g \in (\lambda^{(r-1)}, \lambda^{(r+1)}]$, and $P_\eta$ is the probability distribution of $\eta$.

Re-combining the terms in Equation (41), we have

$$\sum_{r=1}^n \sum_{k=r}^n P_{r,k}^{n-1} P_{\lambda|r,k}^{n-1} \Delta u_{(k)}^{(r)}(\lambda_f, \lambda_g = \lambda^{(r)})$$

$$= -\sum_{r=1}^n \sum_{k=r}^n P_{r,k}^{n-1} P_{\lambda|r,k}^{n-1} \left(\lambda_g - \frac{k-1}{n}\lambda_f - \frac{n+1-k}{n}\lambda^{(k)}\right)(T^{(k)} - T^{(k-1)})$$

$$- \sum_{r=1}^{n-1} \sum_{k=r+1}^n P_{r,k}^{n-1} P_{\lambda|r,k}^{n-1} \left(\sum_{j=r}^{k-1}(\lambda_g - \lambda^{(j+1)})(T^{(j)} - T^{(j-1)})\right.$$

$$\left.- \sum_{j=r}^{k-1}\frac{n-j+1}{n}(\lambda^{(j)} - \lambda^{(j+1)})(T^{(j)} - T^{(j-1)})\right) \leq 0 \tag{42}$$

When $\lambda_f > \lambda_g$,

$$E_{\lambda,\mu,\eta}\left(\Delta u_{(k)}^{(r)}(\lambda_f, \lambda_g)\right) = \int_\lambda \int_\eta \sum_{r=1}^n \sum_{k=1}^r P_{r,k}^{n-1} P_{\lambda|r,k}^{n-1} P_\eta \Delta u_{(k)}^{(r)}(\lambda_f, \lambda_g) d\lambda d\eta \tag{43}$$

Since $\lambda_i$ follows a uniform distribution, $\lambda_i \sim U(\lambda_l, \lambda_u), i \in A$, we have
$$P_{r,k}^{n-1} = P\big(\lambda_f \in (\lambda^{(k-1)}, \lambda^{(k)}], \lambda_g \in (\lambda^{(r-1)}, \lambda^{(r+1)}]\big)$$
$$= \binom{n-1}{k-1}\binom{n-k}{r-k}\left(\frac{\lambda_u - \lambda_f}{\lambda_u - \lambda_l}\right)^{k-1}\left(\frac{\lambda_f - \lambda_g}{\lambda_u - \lambda_l}\right)^{r-k}\left(\frac{\lambda_g - \lambda_l}{\lambda_u - \lambda_l}\right)^{n-r} \quad (44)$$

Re-combining the terms in Equation (43), we have
$$\sum_{r=1}^{n}\sum_{k=1}^{r} P_{r,k}^{n-1} P_{\lambda|r,k}^{n-1} \Delta u_{(k)}^{(r)}(\lambda_f, \lambda_g = \lambda^{(r)})$$
$$= \sum_{r=1}^{n}\sum_{k=1}^{r} P_{r,k}^{n-1} P_{\lambda|r,k}^{n-1} \frac{k-1}{n}(\lambda_f - \lambda^{(k)})(T^{(k)} - T^{(k-1)})$$
$$- \sum_{r=1}^{n}\sum_{k=1}^{r-1} P_{r,k}^{n-1} P_{\lambda|r,k}^{n-1} \left(\sum_{j=k}^{r-1}(\lambda^{(j)} - \lambda_g)(T^{(j+1)} - T^{(j)}) + \sum_{j=k}^{r-1}\frac{n-j}{n}(\lambda^{(j)} - \lambda^{(j+1)})(T^{(j+1)} - T^{(j)})\right)$$
$$= \sum_{r=1}^{n}\sum_{k=1}^{1} P_{r,k}^{n-1} P_{\lambda|r,k}^{n-1} \frac{k-1}{n}(\lambda_f - \lambda^{(k)})(T^{(k)} - T^{(k-1)})$$
$$+ \sum_{r=2}^{n}\sum_{k=2}^{r}\bigg[P_{r,k}^{n-1} P_{\lambda|r,k}^{n-1} \frac{k-1}{n}(\lambda_f - \lambda^{(k)})(T^{(k)} - T^{(k-1)})$$
$$- P_{r,k-1}^{n-1} P_{\lambda|r,k-1}^{n-1}\left(\sum_{j=k-1}^{r-1}(\lambda^{(j)} - \lambda_g)(T^{(j+1)} - T^{(j)}) + \sum_{j=k-1}^{r-1}\frac{n-j}{n}(\lambda^{(j)} - \lambda^{(j+1)})(T^{(j+1)} - T^{(j)})\right)\bigg]$$

From Equation (44), $P_{r,k}^{n-1} = \frac{r-k+1}{k-1}\frac{\lambda_u - \lambda_f}{\lambda_f - \lambda_g} P_{r,k-1}^{n-1}$. Therefore, we have
$$\sum_{r=1}^{n}\sum_{k=1}^{r} P_{r,k}^{n-1} P_{\lambda|r,k}^{n-1} \Delta u_{(k)}^{(r)}(\lambda_f, \lambda_g)$$
$$= \sum_{r=2}^{n}\sum_{k=2}^{r} P_{r,k-1}^{n-1}\bigg[P_{\lambda|r,k}^{n-1} \frac{r-k+1}{k-1}\frac{\lambda_u - \lambda_f}{\lambda_f - \lambda_g}\frac{k-1}{n}(\lambda_f - \lambda^{(k)})(T^{(k)} - T^{(k-1)})$$
$$- P_{\lambda|r,k-1}^{n-1}\left(\sum_{j=k-1}^{r-1}(\lambda^{(j)} - \lambda_g)(T^{(j+1)} - T^{(j)}) + \sum_{j=k-1}^{r-1}\frac{n-j}{n}(\lambda^{(j)} - \lambda^{(j+1)})(T^{(j+1)} - T^{(j)})\right)\bigg]$$
$$\leq \sum_{r=2}^{n}\sum_{k=2}^{r} P_{r,k-1}^{n-1}\bigg[P_{\lambda|r,k}^{n-1} \frac{r-k+1}{k-1}\frac{\lambda_u - \lambda_f}{\lambda_f - \lambda_g}\frac{k-1}{n}(\lambda_f - \lambda^{(k)})(T^{(k)} - T^{(k-1)})$$
$$- P_{\lambda|r,k-1}^{n-1}\left((\lambda^{(k-1)} - \lambda_g)(T^{(k)} - T^{(k-1)}) + \frac{n-k+1}{n}(\lambda^{(k-1)} - \lambda^{(k)})(T^{(k)} - T^{(k-1)})\right)\bigg]$$

Note that $\lambda_i$ follows a uniform distribution, $\lambda_i \sim U(\lambda_l, \lambda_u), i \in A$,
$$\int_\lambda P_{\lambda|r,k}^{n-1}(\lambda_f - \lambda^{(k)})d\lambda = \frac{\lambda_f - \lambda_g}{r-k+1};$$
$$\int_\lambda P_{\lambda|r,k-1}^{n-1}(\lambda^{(k-1)} - \lambda_g)d\lambda = \frac{(r-k+1)(\lambda_f - \lambda_g)}{r-k+2};$$
$$\int_\lambda P_{\lambda|r,k-1}^{n-1}(\lambda^{(k-1)} - \lambda^{(k)})d\lambda = \frac{\lambda_f - \lambda_g}{r-k+2}.$$

Therefore, we have
$$\int_\lambda \sum_{r=1}^{n}\sum_{k=1}^{r} P_{r,k}^{n-1} P_{\lambda|r,k}^{n-1} \Delta u_{(k)}^{(r)}(\lambda_f, \lambda_g) d\lambda$$
$$\leq \sum_{r=2}^{n}\sum_{k=2}^{r} P_{r,k-1}^{n-1}(\lambda_f - \lambda_g)(T^{(k)} - T^{(k-1)})\left(\frac{1}{n}\frac{\lambda_u - \lambda_f}{\lambda_f - \lambda_g} + \frac{1}{n}\frac{k-1}{r-k+2} - 1\right) \quad (45)$$

$$\leq \left[\sum_{r=2}^{n}\sum_{k=2}^{r} P_{r,k-1}^{n-1}(\lambda_f - \lambda_g)\left(\frac{1}{n}\frac{\lambda_u - \lambda_f}{\lambda_f - \lambda_g} + \frac{1}{n}\frac{k-1}{r-k+2} - 1\right)\right] \cdot \left[\sum_{r=2}^{n}\sum_{k=2}^{r}\left(T^{(k)} - T^{(k-1)}\right)\right]$$

$$= (\lambda_f - \lambda_g)\left(\frac{1}{n}\frac{\lambda_u - \lambda_f}{\lambda_f - \lambda_g} + \frac{1}{n}\sum_{r=2}^{n}\sum_{k=2}^{r}\frac{k-1}{r-k+2}P_{r,k-1}^{n-1} - 1\right)\sum_{r=2}^{n}\left(T^{(r)} - T^{(1)}\right)$$

Note that

$$\sum_{r=2}^{n}\sum_{k=2}^{r}\frac{k-1}{r-k+2}P_{r,k-1}^{n-1}$$
$$= \sum_{r=2}^{n}\sum_{k=2}^{r}\frac{k-1}{r-k+2}\frac{(n-1)!}{(k-2)!(r-k+1)!(n-r)!}\frac{(\lambda_u - \lambda_f)^{k-2}(\lambda_f - \lambda_g)^{r-k+1}(\lambda_g - \lambda_l)^{n-r}}{(\lambda_u - \lambda_l)^{n-1}} \quad (46)$$
$$= \frac{\lambda_u - \lambda_l}{n(\lambda_f - \lambda_g)}\sum_{r=2}^{n}\sum_{k=2}^{r}(k-1)P_{r,k-1}^{n}$$

where $P_{r,k-1}^{n}$ is the probability that $\lambda_f$ is larger than or equal to $\lambda^{(k)}$ but smaller than $\lambda^{(k-1)}$ while $\lambda_g$ is larger than or equal to $\lambda^{(r+1)}$ but smaller than $\lambda^{(r-1)}$ when there are $n$ other agents. Note that $\sum_{r=1}^{n}\sum_{k=2}^{r+1}(k-1)P_{r,k-1}^{n}$ can be interpreted as the expected order of agent with $\lambda_f$ among all $n$ agents.

$$\sum_{r=2}^{n}\sum_{k=2}^{r}(k-1)P_{r,k-1}^{n} \leq \sum_{r=1}^{n}\sum_{k=2}^{r+1}(k-1)P_{r,k-1}^{n} = \frac{\lambda_u - \lambda_f}{\lambda_u - \lambda_l}n \quad (47)$$

Combining Equations (46) and (47), we have

$$\sum_{r=2}^{n}\sum_{k=2}^{r}\frac{k-1}{r-k+2}P_{r,k-1}^{n-1} \leq \frac{\lambda_u - \lambda_f}{\lambda_f - \lambda_g} \quad (48)$$

Inserting Equation (48) back into Equation (45), we have

$$\int_{\lambda}\sum_{r=1}^{n}\sum_{k=1}^{r} P_{r,k}^{n-1} P_{\lambda|r,k}^{n-1} \Delta u_{(k)}^{(r)}(\lambda_f, \lambda_g) d\lambda$$
$$\leq (\lambda_f - \lambda_g)\left(\frac{1}{n}\frac{\lambda_u - \lambda_f}{\lambda_f - \lambda_g} + \frac{1}{n}\sum_{r=2}^{n}\sum_{k=2}^{r}\frac{k-1}{r-k+2}P_{r,k-1}^{n-1} - 1\right)\sum_{r=2}^{n}\left(T^{(r)} - T^{(1)}\right) \quad (49)$$
$$\leq \left[\frac{2}{n}(\lambda_u - \lambda_f) - (\lambda_f - \lambda_g)\right]\left(T^{(r)} - T^{(1)}\right)$$
$$\leq \left[\frac{2}{n}(\lambda_u - \lambda_l) - (\lambda_f - \lambda_g)\right]\left(T^{(r)} - T^{(1)}\right)$$

Therefore, when $\lambda_f - \lambda_g \geq \frac{2}{n}(\lambda_u - \lambda_l)$,

$$E_{\lambda,\mu,\eta}\left(\Delta u_{(k)}^{(r)}(\lambda_f, \lambda_g)\right) = \int_{\lambda}\int_{\eta}\sum_{r=1}^{n}\sum_{k=1}^{r} P_{r,k}^{n-1} P_{\lambda|r,k}^{n-1} P_{\eta} \Delta u_{(k)}^{(r)}(\lambda_f, \lambda_g) d\lambda d\eta$$
$$\leq \left[\frac{2}{n}(\lambda_u - \lambda_l) - (\lambda_f - \lambda_g)\right]\int_{\eta}\sum_{r=2}^{n}\left(T^{(r)} - T^{(1)}\right) d\eta$$
$$\leq 0. \qquad \square$$